\documentclass[a4paper,fleqn,usenatbib]{mnras}

\usepackage{newtxtext,newtxmath}

\usepackage[T1]{fontenc}
\usepackage{ae,aecompl}

\usepackage{graphicx}	
\usepackage{amsmath}	
\usepackage{amssymb}	
\usepackage{subfig}

\newcommand{\oii}[1]{[\textsc{O\,ii}]#1}

\newcommand{\oiii}[1]{[\textsc{O\,iii}]#1}  
\newcommand{\nii}[1]{[\textsc{N\,ii}]#1}

\newcommand{\ha}{\ensuremath{\mathrm{H}\alpha}} 
\newcommand{\hb}{\ensuremath{\mathrm{H}\beta}} 
 
\newcommand{\hii}{H\,\textsc{ii}}

\newcommand{\Mstar}{\ensuremath{M_*}} 
 
\newcommand{\Msun}{\ensuremath{\mathrm{M}_\odot}} 
\newcommand{\Zsun}{\ensuremath{\mathrm{Z}_\odot}} 
\newcommand{\Z}{\ensuremath{\mathrm{Z}}} 
 
\newcommand{\logU}{$\log U$}

\newcommand{\SFRUV}{\ensuremath{\mathrm{SFR}_{\mathrm{UV}}}}
\newcommand{\SFRHa}{\ensuremath{\mathrm{SFR}_{\mathrm{H}\alpha}}}
\newcommand{\LUV}{\ensuremath{\mathrm{L}_{\mathrm{UV}}}}
\newcommand{\LHa}{\ensuremath{\mathrm{L}_{\mathrm{H}\alpha}}}
\newcommand{\SFR}{\ensuremath{\mathrm{SFR}}}
\newcommand{\SFRs}{\ensuremath{\mathrm{SFRs}}}
\newcommand{\SED}{\ensuremath{\mathrm{SED}}}

\newcommand{\Q}{\ensuremath{\mathrm{Q}}}

\newcommand{\rin}{\ensuremath{\mathrm{r}_{in}}} 
\newcommand{\RS}{\ensuremath{R_{\mathrm{S}}}} 
\newcommand{\nh}{\ensuremath{n_\mathrm{H}}}    
\newcommand{\ab}{\ensuremath{\mathrm{\alpha}_{B}}}

\newcommand{\rtt}{R{\small 23}}
\newcommand{\otnt}{O{\small 3}N{\small 2}}
\newcommand{\ntot}{N{\small 2}O{\small 2}}
\newcommand{\ott}{O{\small 32}}
\newcommand{\Te}{\ensuremath{T}{\small e}}

\title[Metallicity calibrations of galaxies with low SFR]{Metallicity calibrations of low star-forming galaxies: the influence of a stochastic IMF}

\author[M. Paalvast and J. Brinchmann]{
Mieke Paalvast$^{1}$\thanks{E-mail: paalvast@strw.leidenuniv.nl}
and Jarle Brinchmann$^{1,2}$
\\
$^{1}$Leiden Observatory, Leiden University, PO Box 9513, NL-2300 RA Leiden, the Netherlands \\
$^{2}$Instituto de Astrof{\'i}sica e Ci{\^e}ncias do Espaço, Universidade do Porto, CAUP, Rua das Estrelas, PT4150-762 Porto, Portugal\\
}

\date{Accepted 2017 May 19. Received 2017 May 16; in original form 2017 March 6.}

\pubyear{2017}

\begin{document}
\label{firstpage}
\pagerange{\pageref{firstpage}--\pageref{lastpage}}
\maketitle

\begin{abstract}
We present a study of the consequences of an initial mass function that is stochastically sampled on the main emission lines used for gas-phase metallicity estimates in extra-galactic sources. We use the stochastic stellar population code SLUG and the photoionisation code Cloudy to show that the stochastic sampling of the massive end of the mass function can lead to clear variations in the relative production of energetic emission lines such as \oiii\ relative to that of Balmer lines.  We use this to study the impact on the \Te, \ntot, \rtt\ and \otnt\ metallicity calibrators. We find that stochastic sampling of the IMF leads to a systematic over-estimate of O/H in galaxies with low star formation rates ($\le 10^{-3}$ \Msun /yr) when using the \ntot, \rtt\ and \otnt\ strong-line methods, and an under-estimate when using the \Te\ method on galaxies of sub-solar metallicity. We point out that while the  \SFRHa -to-\SFRUV\ ratio can be used to identify systems where the initial mass function might be insufficiently sampled, it does not provide sufficient information to fully correct the metallicity calibrations at low star formation rates. Care must therefore be given in the choice of metallicity indicators in such systems, with the \ntot\ indicator proving most robust of those tested by us, with a bias of 0.08 dex for models with \SFR\ = 10$^{-4}$ \Msun /yr and solar metallicity.
\end{abstract}

\begin{keywords}
Galaxies: dwarf -- Galaxies: abundances -- Galaxies: star formation -- ISM: H II region -- Galaxies: evolution
\end{keywords}




\section{Introduction}
Understanding the chemical enrichment of galaxies furthers our understanding of galaxy evolution. Elements heavier than hydrogen and helium are produced in stellar cores or during super nova explosions, which are both linked to the star formation of a galaxy. Therefore, measuring metallicities provides an insight into the history of star formation. Together with stellar mass (\Mstar) and current star formation rate (\SFR), determinations of the metallicity are essential to constrain models of chemical enrichment in galaxies and can with some simplifications be used to place constraints on the overall evolution of galaxies (e.g \citealt{2012MNRAS.421...98D,2017MNRAS.tmp..110D,2013ApJ...772..119L, 2015ApJ...808..129L,  2016MNRAS.456.2140M} ).

Nebular emission lines from \hii \space regions provide most of the information we have on the chemical abundances in the ionised gas in distant galaxies, see for example \citet{2004ApJ...613..898T, 2002ApJ...581.1019G, 2015MNRAS.451.2251I, 2014ApJ...789L..40W}. For massive galaxies this is a fairly well-established process, and the resulting gas-phase metallicities have been related to other physical properties, such as the \SFR\ and \Mstar, although some caveats remain \citep{2008ApJ...681.1183K, 2010MNRAS.408.2115M, 2012MNRAS.422..215Y, 2013ApJ...764..178L, 2014MNRAS.443.2695S, 2015AJ....149...79D}. Relationships such as the \Mstar\ - \SFR\ \citep{2004MNRAS.351.1151B, 2007ApJ...660L..43N, 2012ApJ...754L..29W}, and \Mstar\ - metallicity (\Z) \citep{2004ApJ...613..898T, 2014ApJ...791..130Z} relationships are now well characterised for more massive galaxies and have yielded important insights into the evolution of galaxies: the tightness of the \Mstar\ - \SFR\ relationship, for instance, argues for a equilibrium model of galaxy evolution \citep{2013ApJ...772..119L}.

The main focus of the studies above has been on the more massive galaxies. However, expanding our understanding of the physical processes of small systems is of interest, because theories about galaxy evolution predict that low-mass and low star-forming galaxies might be analogs to high redshift galaxies and progenitors of higher mass systems. Moreover, they provide insight in feedback processes since small systems should be most affected by them. Although the relation between star formation properties suggest that the processes in galaxies with low \SFR, \Mstar\ and \Z\ are related in the same way for low mass systems, a deviating slope has been derived for the low-mass end of the \Mstar\ - \SFR\ \citep{2014ApJ...795..104W} and the \Mstar\ - \Z\ \citep{2012ApJ...750..120Z} relation. However, empirical evidence from a large sample of galaxies for this is missing. 

One of the challenges of determining the \SFR\ and the \Z\ of low-mass galaxies is the assumption in traditional methods that the initial mass function (IMF) of the stellar population in the galaxy is fully populated. In the work of \citet{2009ApJ...706..599L,2009ApJ...695..765M} amongst others, it is shown that the \SFR\ that is derived from the \ha\ luminosity is inconsistent with the \SFR\ estimated from UV continuum light for galaxies with a low \SFR. Because the \ha\ line is due to stars with higher masses than those dominating the UV continuum, this is an indication that the stellar mass distribution in their galaxies is different from that expected from a fully populated normal IMF.

The logical interpretation of this inconsistency is that there is variation in the sampling of the massive end of the IMF. Different ways to achieve this were examined by \citet{2011ApJ...741L..26F} who find that both an integrated galactic IMF (IGIMF) and a stochastically sampled IMF can cause low values of \LHa/\LUV. For both scenarios, stars are assumed to be forming in embedded star clusters. The IGIMF is a modified IMF in which the maximum stellar mass is a function the mass of its birth cluster, whereas in a stochastically sampled cluster the maximum stellar mass varies more. \citet{2000ApJ...539..342E} finds that the maximum mass of a star in a coeval population is related to the total mass of the population. For example, a population of at least $10^4$ \Msun\ is needed for the formation of one star with mass 120 \Msun . In low star-forming galaxies, the mass of birth clouds is often lower than this limit and may lead to truncation of the IMF \citep{2004A&A...413..145C}. However,  \citet{2011ApJ...741L..26F} argue that the observationally derived discrepancy between \SFRHa\ and \SFRUV\ can best be explained by a scenario where the stellar masses in a coeval stellar population are distributed in a random way, and where the distribution of stars varies for different populations of the same total mass. This is referred to as 'stochastic sampling'. 

There are other explanations for the variation in \LHa/\LUV, besides a stochastically sampled IMF. As we mentioned before, the IGIMF \citep{2005ApJ...625..754W, 2007ApJ...671.1550P} can explain the difference in the \SFR\ results as shown by \citet{2009ApJ...706..599L}. Moreover, \citet{2016ApJ...833...37G} conclude that low-mass galaxies have bursty star formation histories, which can be responsible for the difference between  \SFRHa\ and \SFRUV \citep{2014ApJ...789..147W,2009ApJ...706.1527B}. A number of other results have also argued for some variation in the IMF. At the upper end of the mass spectrum of galaxies, \citet{2010Natur.468..940V} have demonstrated that elliptical galaxies show evidence of an IMF systematically different from that commonly inferred for spiral disks. A different slope of the IMF could be responsible for the discrepancy, and for example \citet{2015MNRAS.447..618B} found a steeper slope for a blue compact dwarf galaxy.  Other examples to explain the variation in \SFRHa/\SFRUV\ can be found in a possible leakage of ionising photons and uncertainties in dust corrections. However, for this paper we will only consider a stochastically sampled IMF, but we will discuss the consequences of a different IMF choice in the discussion.
 
A stochastically sampled IMF potentially affects any physical property inference that rely on strong emission lines. For this paper we investigate how a stochastic distribution of stellar masses influences the  chemical abundance determinations of galaxies with a low \SFR. We will use the "Stochastically Lighting Up Galaxies" (SLUG) code \citep{2012ApJ...745..145D, 2014MNRAS.444.3275D, 2015MNRAS.452.1447K} to model galaxies with low star formation. We combine this with nebular \hii\ region simulations from Cloudy \citep{2013RMxAA..49..137F} to analyse the influence of varying massive star distributions on four different abundance determination methods. 

In what follows, we start with a description of our stellar and nebular models. This is followed by a comparison of our results with SDSS DR7 data, using diagnostic emission-line diagrams to demonstrate that our nominal model can reproduce observational data. Then we focus on measuring the chemical abundances of our models using the direct \Te\ method, as well as three commonly used calibrators with the aim of investigating how much the results are affected by the stochastic IMF.  This is followed by a discussion of the physical interpretation of our results and of the detectability of a stochastic IMF, which can be used for future improvements of metallicity calibration of low star-forming galaxies. We finish with a discussion on how other variations stellar mass distributions, beside stochastic sampling, would influence metallicity measurements. 


\section{Modelling}

\subsection{Stellar emission}
The stellar initial mass function describes the birth mass distribution of stars within a star forming region \citep{1955ApJ...121..161S, 2001MNRAS.322..231K, 2003PASP..115..763C}. It is usually assumed to be constant for all stellar populations regardless of the stellar properties and the formation time of the population \citep{2010ARA&A..48..339B}. 
Here we use the approach  to model galaxies with a low \SFR\, where star formation in exclusively happening within clusters, as described by \citet{2011ApJ...741L..26F}, using the code SLUG version 2\footnote{\url{https://bitbucket.org/krumholz/slug2}} 
\citep{2012ApJ...745..145D, 2014MNRAS.444.3275D, 2015MNRAS.452.1447K}. In this scenario, the mass of stellar clusters is distributed by a probability distribution functioned called the cluster mass function (CMF). \citet{2012ApJ...745..145D} argue that according to observations \citep{1999ApJ...527L..81Z, 2003ARA&A..41...57L,2009ApJ...704..453F, 2010ApJ...711.1263C} and theory \citep{2010ApJ...710L.142F} the CMF is best described by a power-law with index 2, and we will adopt this here. Thus the CMF is given by
\begin{equation}
dN/dM \propto M^{-2}
\end{equation} 
with cluster mass, $M$, within the range $10^2$ - $10^7$ \Msun. We adopt a Salpeter IMF \citep{1955ApJ...121..161S}, resulting in a distribution of stellar mass $m$ given by,  
\begin{equation}
dN/dm \propto m^{-2.35}
\end{equation} 
with lower and upper mass cut-off of  0.08 \Msun\ and 100 \Msun. After time $t_\mathrm{clus}$ , clusters will eventually disrupt following the cluster lifetime function (CLF) between 1 Myr and 1 Gyr: 
\begin{equation}
dN/d t_\mathrm{clus} \propto t_\mathrm{clus}^{-1.9}.
\end{equation} 
We assume that all stars form in clusters, and that once clusters are disrupted, their stellar radiation still contributes to the total light of the galaxy. 

\begin{figure}
    \includegraphics[width=\columnwidth]{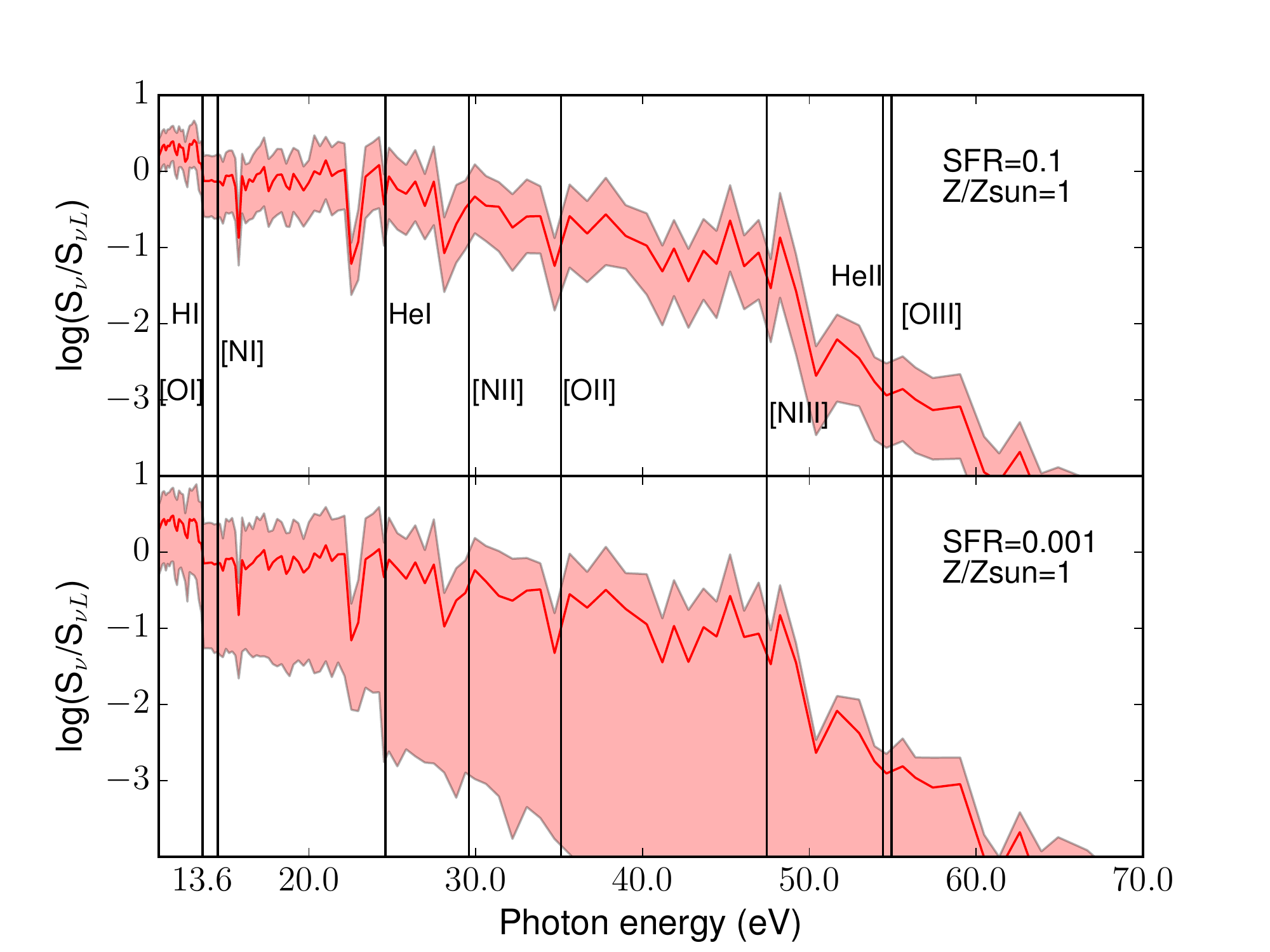}
    \caption{The mean and the 2 $\sigma$ spread of the spectral energy distributions of our stellar models from SLUG with solar metallicity and \SFR = 0.1 \Msun /yr (top panel) and \SFR = 0.001 \Msun /yr (bottom panel), given in units of the luminosity per unit frequency at the Lyman limit. Each panel consist of the result of 100 stochastic galactic models for which we adopted a solar metallicity. The red solid lines show the average SED with the red shaded region encompassing the region where 90 $\%$ of the spectra fall. The variance in the FUV increases strongly with decreasing SFR.  The black lines represent the ionisation energies of various ions, required for the indicated families of lines.}
    \label{fig:slug_sed}
\end{figure}

In situations of low \SFR \space ($\lesssim$ 0.01 \Msun/yr), star formation in clusters leads to variations in the number of massive stars for a galaxy, resulting in variations in the  far-extreme UV (where $E$ > 13.6 eV) of the spectral energy distribution (\SED) as presented in Figure \ref{fig:slug_sed}. In this figure, the top panel shows the mean \SED \space and the 90 \% scatter of 100 models with \SFR\ = 0.1 \Msun/yr. As a comparison, in the bottom panel we present a similar plot with \SFR\ = 0.001 \Msun/yr. The \SED s are normalised at the Lyman limit at 13.6 eV, and ionisation energies for several transitions that are important for determining physical properties of galaxies are visualised by the black lines. Note the large spread in the ionising continuum between models with the same SFR. We will see later that this has clear consequences for emission-line production. 

\subsection{Transmission through the ISM}
To calculate the nebular emission line luminosities of our models we use the SLUG interface with the Cloudy code v13.03, last described by \citet{2013RMxAA..49..137F}.  We normalize the energy distribution of the incident light to a fixed amount of hydrogen ionising photons. The galaxy is assumed to be point like and the nebula is described as spherical layers centered on the ionising source. The geometry of the \hii \space model is defined by the ratio between the innermost layer \rin \space and the radius of the Str{\"o}mgren sphere \RS, defined as \citep{1989agna.book.....O}:
\begin{equation}
\RS^3 = 3Q/(4\pi \nh^2 \epsilon \ab).
\label{eqn:rstrom}
\end{equation} 
Here, $Q$ is the amount of ionising photons per second, that is determined by the SLUG models, \nh \space is the hydrogen density, $\epsilon$ is the filling factor and \ab \space is the Case-B hydrogen recombination coefficient. The influence of the ionising photons is defined by the ionisation parameter that is equal to the rate of ionising photons from the source over the density at a distance $r$ from the source, given by
\begin{equation}
U(r) = Q/(4 \pi r^2 \nh c).
\end{equation} 
To determine the geometry of the system, we assume an inner radius of the nebula to be much smaller than the Str{\"o}mgren radius, \rin $\ll$ \RS, so $U(r)$ is strongly dependent on r, and the total thickness of the \hii \space region is similar to \RS\ (\citealt{2001MNRAS.323..887C}, see also \citealt{2016MNRAS.462.1757G}). This leads to a spherical geometry, for which the volume-averaged ionisation parameter equals
\begin{equation}
\langle U(r) \rangle \approx 3Q/(4 \pi \RS^2 \nh c) = 3U(\RS).
\label{eqn:uavg}
\end{equation}
After substituting equation \ref{eqn:rstrom} into equation \ref{eqn:uavg}, this leads to the relation:
\begin{equation}
\langle U(r) \rangle \approx \frac{\ab^{2/3}}{c} (\frac{3Q \epsilon^2 \nh}{4 \pi})^{1/3}.
\label{eqn:ufin}
\end{equation}
To distinguish the effect of the ionisation parameter and the distribution of ionising photons per energy, we set the amount of ionising hydrogen photons, as given by 
\begin{equation}
U (H_0) = \frac{Q (H)}{4 \pi r_{0}^{2} n (H) c} = \frac{R_{\star}^2}{ r_{0}^{2} n (H) c} \int_{\nu_{1}}^{\nu_{2}} \! \frac{\pi F_\nu}{h \nu} \, \mathrm{d}\nu .
\label{eqn:U}
\end{equation} 
The choice of $\nu_1$ is straightforward and is taken to be 1.0 Ry, but the value for $\nu_2$ merits some discussion. In most studies, and the default in Cloudy, $\nu_2=\infty$, which sums up all photons with energy sufficient to ionise hydrogen. However, as the photo-ionisation cross-section of hydrogen is a strong function of energy, photons close to 1 Ry are the most important for hydrogen ionisation and therefore also for the \ha\ luminosity of a galaxy. For a fully sampled IMF, the total number of ionising photons is directly related to the hydrogen ionising photons and thus $\nu_2=\infty$ can be naturally taken to be the total number of photons with energy above 1 Ry, $Q(E>1\,\mathrm{Ry})$. However, for a stochastically sampled IMF the situation is somewhat more subtle. The same number of photons with $1.0 \space \mathrm{Ry} < E <1.5 \space \mathrm{Ry}$, and hence the same \ha\ flux, can be produced by stellar population with significantly different number of photons with $E>1\,\mathrm{Ry}$. If we were to use the standard definition of $U$, we would end up with results that can differ significantly from the standard results in the literature, such as \citet{2001MNRAS.323..887C, 2013ApJS..208...10D, 2016MNRAS.462.1757G},  particularly at low \SFRs. To mitigate against this and ensure that our $U$ values can be compared to other work in the literature, we define the $U$ parameter by integrating over a smaller range in frequency, e.g. we fix  $\nu_1$ and $\nu_2$ to 1.0 and 1.5 respectively and we refer to this $U$ as  $U (H_0)$.  We set $U (H_0)$ in such a way that for our non-stochastic models, the extrapolated total amount of ionising photons equals that of the models in other studies with $\log U (tot)$ = [-4, -3, -2]. 

\begin{figure}
  \includegraphics[width=\columnwidth]{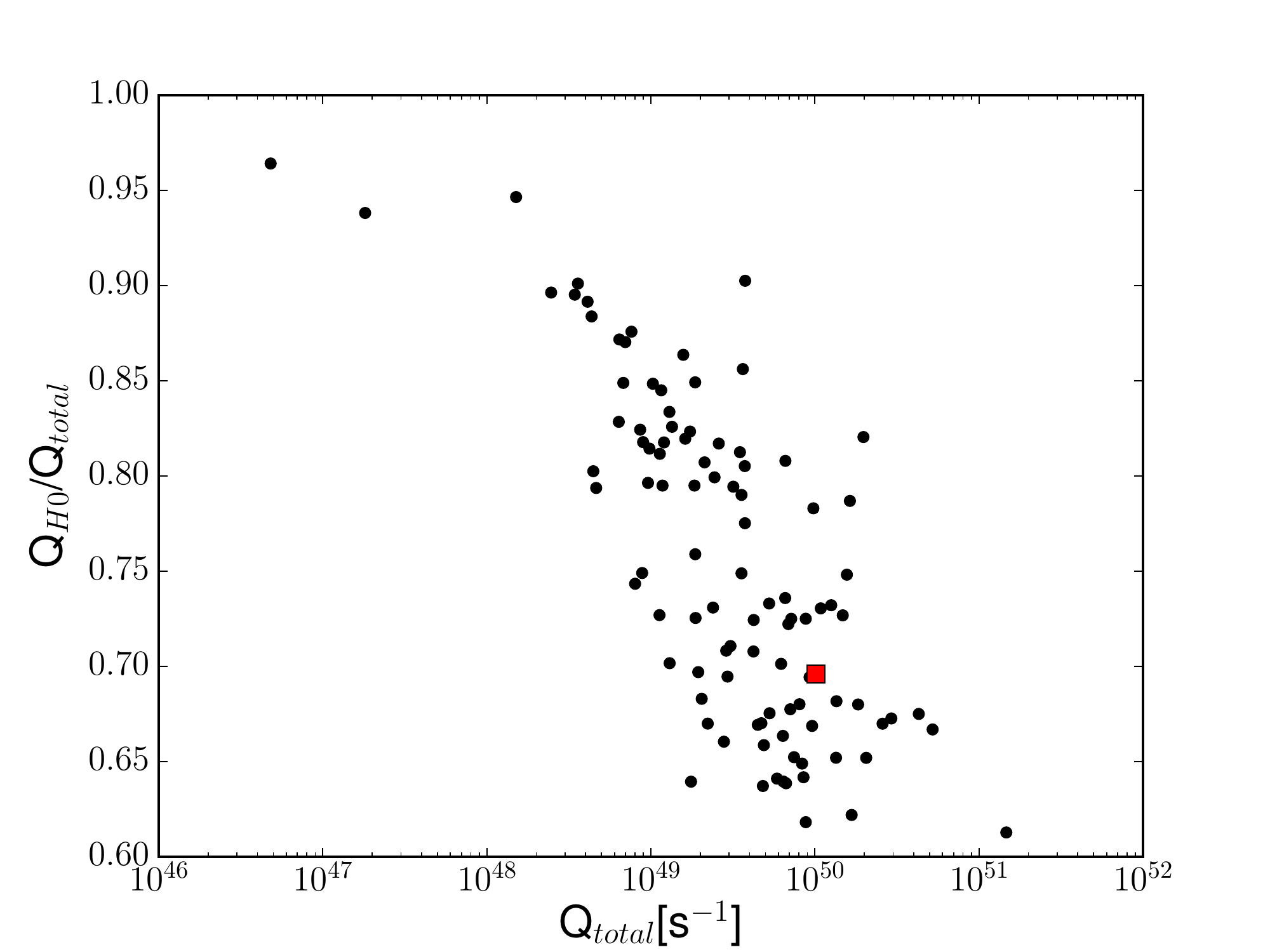}
\caption{The ratio of hydrogen ionising photons over the total ionising photons versus the total amount of photons with $E>1\,\mathrm{Ry}$, of models with \SFR\ = 0.001 \Msun/yr and solar metallicity with a stochastically sampled IMF (black dots) and a non-stochastic model (red square). For the hydrogen ionising photons we have taken the interval from 1.0 to 1.5 Ry. }
\label{fig:Qratio}
\end{figure}

We show the difference between the amount of hydrogen ionising photons ($Q_{H0}$) and the total amount of ionising photons ($Q_{total}$) in Figure \ref{fig:Qratio}. Here we present $Q_{H0}$/$Q_{total}$ versus $Q_{total}$ of models with \SFR\ = 0.001 \Msun/yr and solar metallicity. The number of hydrogen ionising photons to the total amount of ionising photons is not identical for the stochastic models (black dots) as for the non-stochastic model in this bin (red square). This leads to a slight offset in the total amount of ionising photons ($Q_{total}$) of our models with a stochastically sampled IMF compared to those emitted by the non-stochastic model. 

Further, we adopt a filling factor $\epsilon$ of 0.1 \citep{2013ApJ...769...94Z} and a hydrogen density $n_{\mathrm{H}}$ of 100 cm$^{-3}$ (identical to that of the 'standard' model of \citealt{2016MNRAS.462.1757G}) . The gas metallicity equals the metallicity of the stars in the SLUG models, where we adopt solar abundances log(O/H)${_\odot}$ following \citet{2016MNRAS.462.1757G}, which is mostly based on \citet{2012MNRAS.427..127B}. Except for helium and nitrogen, we assume that all the elements heavier than hydrogen scale with oxygen for different chemical abundances. For our nitrogen abundances we use the relation from \citet{2004ApJS..153....9G}, given by
\begin{equation}
[\mathrm{N}/\mathrm{H}]= [\mathrm{O}/\mathrm{H}]10^{-1.6} + 10^{(2.33+\log_{10}[\mathrm{O}/\mathrm{H}])}).
\end{equation}
The helium abundances are as described by \citet{2012MNRAS.427..127B}. 

A fraction of the elements will be captured onto dust grains and the actual gas abundances will therefore be lower than described above. We adopt depletion factors from \citet{2013ApJS..208...10D}.  The effect of this on oxygen is particularly relevant to our work here, and it leads to a gas-phase oxygen abundance of  [O/H] = -3.17 whereas the total oxygen abundance is  [O/H] = -3.32. We also use standard dust properties in Cloudy with dust grain geometries from \citet{1977ApJ...217..425M} and scattering properties from \citet{1991eua..coll..341M}. These latter choices are of minor importance for our work. 

\section{Optical emission-line diagnostics}
In this section, we present the emission line properties of our different models. We first explain our model sampling, then we will compare our results with observational data, followed by a comparison between stellar light and emission-line properties, to study the effect of the stochastically sampled IMF. 

\subsection{Model parameters}
We created 48 bins, each consisting of 100 galaxy models, with different values for the metallicity, the ionisation parameter and the star formation rate. A summary is given in Table \ref{tab:slug_cloudy}.  \\
\\
\textbf{Star formation rate}  In order to study the influence of the \SFR\ on emission-line properties of galaxies  with a stochastically sampled IMF, we divided our sample in bins with \SFR\ = [0.0001, 0.001, 0.01, 0.1], as this reasonably spans the range of galaxies studied in nearby samples such as 11HUGS (e.g. \citealt{2009ApJ...706..599L}). However, \SFR s below 0.01 \Msun/yr are rare in magnitude limited surveys. For example, in the SDSS DR7 data sample \citep{2004MNRAS.351.1151B, 2004ApJ...613..898T}, 0.89 $\%$ (1817 out of 203219) of all star-forming galaxies in their sample have a \SFR\ below 0.01 \Msun/yr, 0.19 $\%$ (379) below \SFR\ = 0.001  \Msun/yr and 0.05 $\%$  (95) below \SFR\ = 0.0001  \Msun/yr.\\
\\
\textbf{Stellar metallicity}  We adopted Geneva stellar tracks with mass-loss \citep{2012A&A...541A..41M}. \\
\\
\textbf{Nebular metallicity}  The nebular metallicity (gas + dust) is assumed to be identical to that of the ionising stars.  Although we are aware that there is a possibility that these deviate for star-forming galaxies (e.g. \citealt{2016ApJ...826..159S}), we point out that the metallicity of the ionising stars is of main importance for our study, and this is more likely to be similar to the nebular metallicity than the mean metallicity of all stars. A part of these metallicities is depleted onto dust grains. We adopt a dust-to-metal mass ratio $\xi_d$ = 0.36 (= $\xi_\odot$) \citep{2016MNRAS.462.1757G} and depletion factors from \citet{2004ApJS..153....9G}. \\
\\
\textbf{Ionisation parameter}  We normalized the energy of hydrogen ionising photons to match the total ionising energy  $\log U$ = [-4, -3, -2], for a non-stochastic model, as described in the previous section. 
\begin{table}
	\centering
	
	\begin{tabular}{ccc} 
		\hline
		\hline
		 SFR (\Msun/yr) & [0.0001, 0.001, 0.01, 0.1] \\
		\hline
		Stellar and nebular abundances (\Zsun) &  [0.05, 0.2, 0.4, 1.0] \\
		\hline
		\logU\ (total ionising energy) & [-4, -3, -2] \\
		\hline
		inner radius $R_{\mathrm{in}}$ (pc) & 0.001 \\
		\hline
		hydrogen density (cm$^{-2}$) & 100 \\
		\hline
		dust-to-metal mass ratio ($\xi_d$) & 0.36 \\
		\hline
		\hline
    	\end{tabular}

	\caption{Summary of our model bins as described by the text.}
        \label{tab:slug_cloudy}
\end{table}
\subsection{Comparison with observations}
\begin{figure*}
   \begin{minipage}{1.0cm}
  \vspace{8.0cm}
  \hspace{-9.1cm}
  \rotatebox{90}{ \large{log (\oiii$\lambda$5007/\hb)}}
 \vspace{-8.0cm} 
 \hspace{9.1cm}
  \end{minipage}%
  \vspace{-3.0cm}
         \hspace*{-0.4cm}

     \subfloat{
       \includegraphics[width=0.252\textwidth]{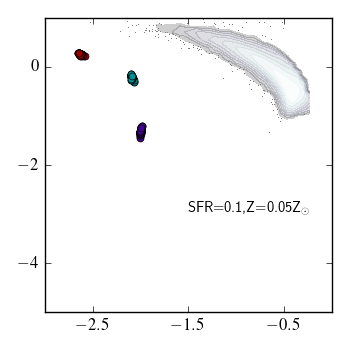}
     }
    \hspace*{-0.4cm}
     \subfloat{
       \includegraphics[width=0.252\textwidth]{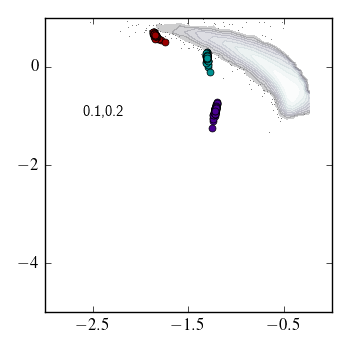}
     }
      \hspace*{-0.4cm}
          \subfloat{
       \includegraphics[width=0.252\textwidth]{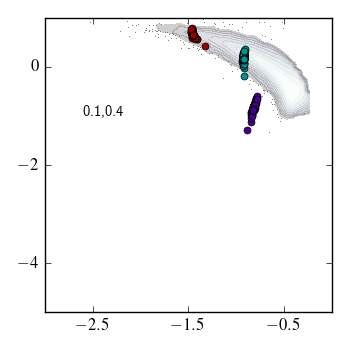}
     }
      \hspace*{-0.4cm}
     \subfloat{
       \includegraphics[width=0.252\textwidth]{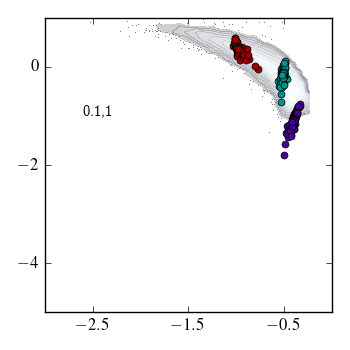}
     }
     \hfill

     \vspace*{-0.5cm}
          \subfloat{
       \includegraphics[width=0.252\textwidth]{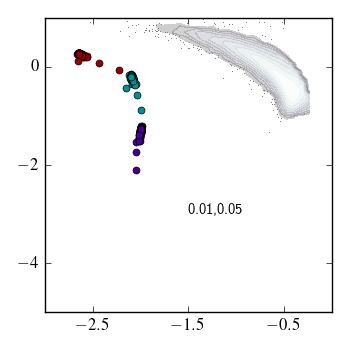}
     }
     \hspace*{-0.4cm}
     \subfloat{
       \includegraphics[width=0.252\textwidth]{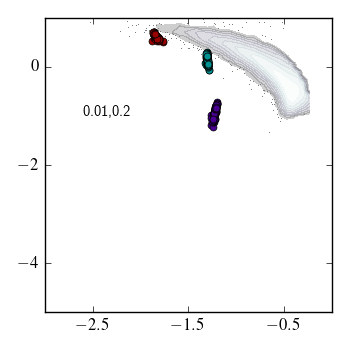}
     }
     \hspace*{-0.4cm}
          \subfloat{
       \includegraphics[width=0.252\textwidth]{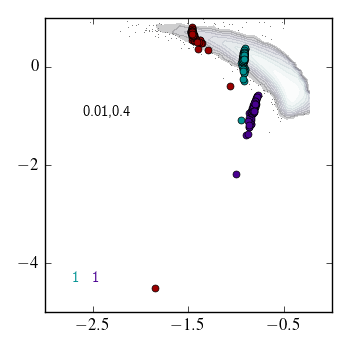}
     }
     \hspace*{-0.4cm}
     \subfloat{
       \includegraphics[width=0.252\textwidth]{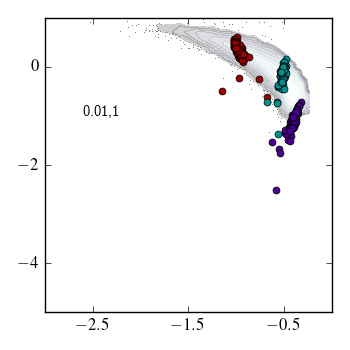}
     }
     \hfill
         \vspace*{-0.5cm}
          \subfloat{
       \includegraphics[width=0.252\textwidth]{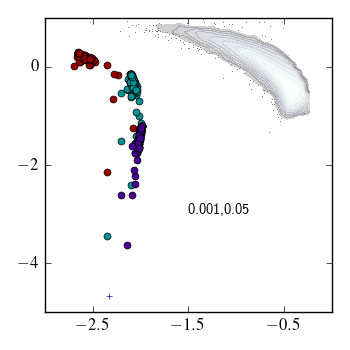}
     }
     \hspace*{-0.4cm}
     \subfloat{
       \includegraphics[width=0.252\textwidth]{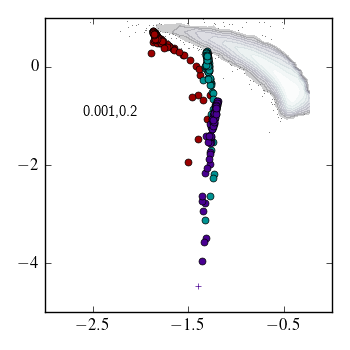}
     }
     \hspace*{-0.4cm}
          \subfloat{
       \includegraphics[width=0.252\textwidth]{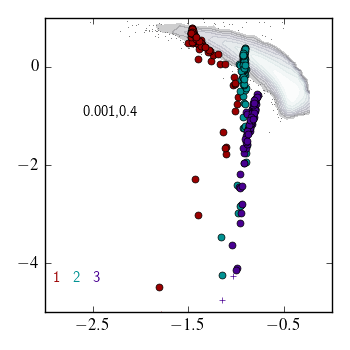}
     }
     \hspace*{-0.4cm}
     \subfloat{
       \includegraphics[width=0.252\textwidth]{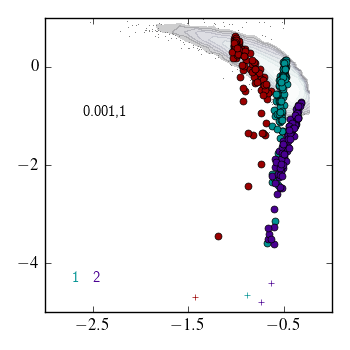}
     }
     \hfill
      \vspace*{-0.5cm}
          \subfloat{
       \includegraphics[width=0.252\textwidth]{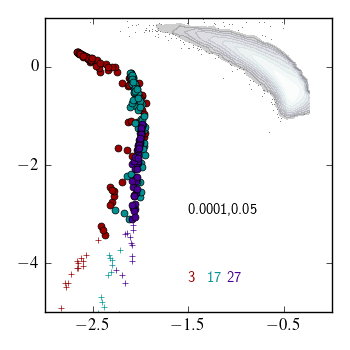}
     }
     \hspace*{-0.4cm}
     \subfloat{
       \includegraphics[width=0.252\textwidth]{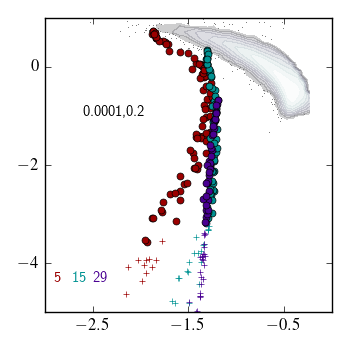}
     }
     \hspace*{-0.4cm}
          \subfloat{
       \includegraphics[width=0.252\textwidth]{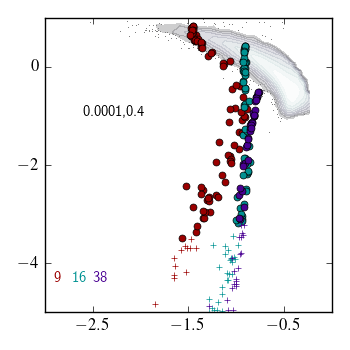}
     }
     \hspace*{-0.4cm}
     \subfloat{
       \includegraphics[width=0.252\textwidth]{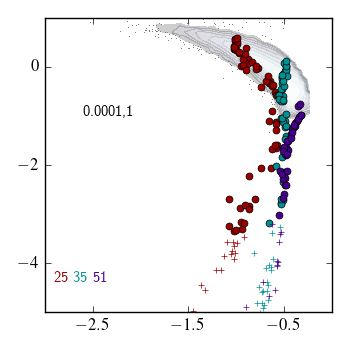}
     }
  \begin{minipage}{1.0cm}
  \vspace{0.5cm}
  \hspace{-10.4cm}
  \rotatebox{0}{ \large{log (\nii$\lambda$6584/\ha)}}
 \vspace{-0.5cm} 
 \hspace{10.4cm}
  \end{minipage}
     \vspace{0.5cm}
     \caption{The behaviour of our models in the \nii/\ha\ vs \oiii/\hb\ diagram. Each panel corresponds to one combination of SFR and metallicity, indicated in the bottom left of each panel. The SFR  decreases from the top to the bottom in steps of a factor of 10 (0.1, 0.01, 0.001) and the metallicity increases left to right (\Z/\Zsun\ = 0.05, 0.2, 0.4, 1.0). The red filled circles are for $\log U (tot)$ = -2, the cyan for $\log U (tot)$ = -3 and the purple for $\log U (tot)$ = -4. The coloured numbers present the amount of models with $\log$\oiii/\hb\ $<$ -5. The gray-scale underneath shows the distribution of star-forming galaxies in the SDSS DR7 and shows the locus of normal, relatively massive, galaxies. At the highest SFR the models trace the locus of the SDSS galaxies, but at lower SFR the scatter, particularly at the lowest U value, increases dramatically and extends into the part of the parameter space where normal galaxies do not fall.}
     \label{fig:lineratio1}
   \end{figure*}

Figure \ref{fig:lineratio1} shows the \oiii{\ensuremath{\lambda}5007}/\hb\ versus \nii{\ensuremath{\lambda}6584}/\ha\ BPT-diagram \citep{1981PASP...93....5B}, which is sensitive to the ionisation parameter at fixed metallicity. The \SFR\ is decreasing from the top to the bottom panel and the chemical abundances are increasing from left to right. The colours indicate different values of $\log U$. 
When the flux of one of the lines is below a detection limit of $10^{-18}$ erg/s/cm$^2$, the result is presented as a cross, above this limit as a dot. We set this detection limit in such a way that it equals the depth that is reached in deep fields with MUSE \citep{2015A&A...575A..75B}. We see here that all models, irrespective of SFR, reach similarly high \oiii/\hb\ values, but at lower \SFR\ the scatter towards low \oiii/\hb\ increases significantly. We compare the models with SDSS DR7 data (grey contours), from which the derivation of the measurements is detailed in \citet{2004MNRAS.351.1151B} and \citet{2004ApJ...613..898T}. We only included the SDSS observations of galaxies with a signal-to-noise of at least 3 in all the lines that are used for this plot . At the highest SFR the models trace the locus of the SDSS galaxies, but at lower SFR the scatter, particularly at the lowest U value, increases dramatically and extends into the part of the parameter space where normal galaxies do not fall. As an example, only 0.013$\%$ (26 out of 203219) of the star forming galaxies in the SDSS sample have a $\log$ \oiii/\hb\ value below -1, with a minimum $\log$ \oiii/\hb\ of -1.33.

\begin{figure*}
   \begin{minipage}{1.0cm}
  \vspace{7.2cm}
  \hspace{-9.1cm}
  \rotatebox{90}{ \large{log (\oiii$\lambda$5007/\oii$\lambda$3727)}}
 \vspace{-7.2cm} 
 \hspace{9.1cm}
  \end{minipage}%
  \vspace{-4.5cm}
         \hspace*{-0.4cm}

     \subfloat{
       \includegraphics[width=0.252\textwidth]{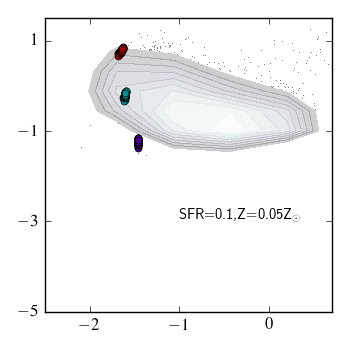}
     }
    \hspace*{-0.4cm}
     \subfloat{
       \includegraphics[width=0.252\textwidth]{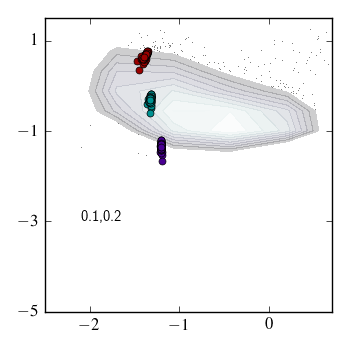}
     }
      \hspace*{-0.4cm}
          \subfloat{
       \includegraphics[width=0.252\textwidth]{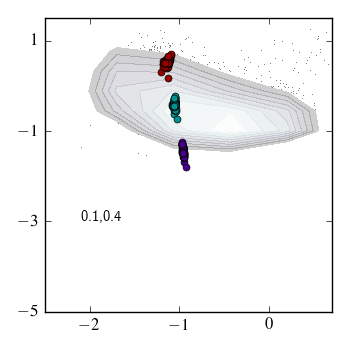}
     }
      \hspace*{-0.4cm}
     \subfloat{
       \includegraphics[width=0.252\textwidth]{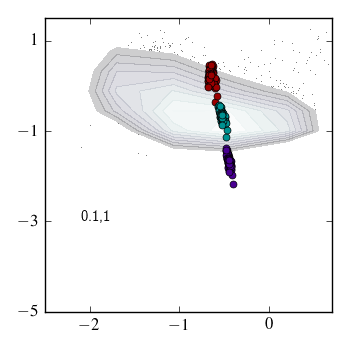}
     }
     \hfill

     \vspace*{-0.5cm}
          \subfloat{
       \includegraphics[width=0.252\textwidth]{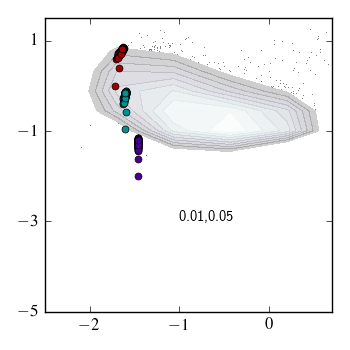}
     }
     \hspace*{-0.4cm}
     \subfloat{
       \includegraphics[width=0.252\textwidth]{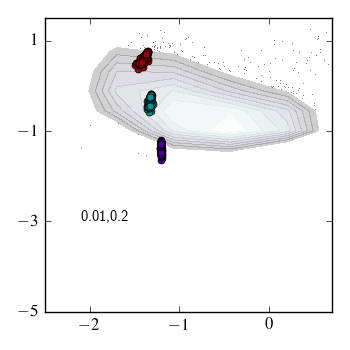}
     }
     \hspace*{-0.4cm}
          \subfloat{
       \includegraphics[width=0.252\textwidth]{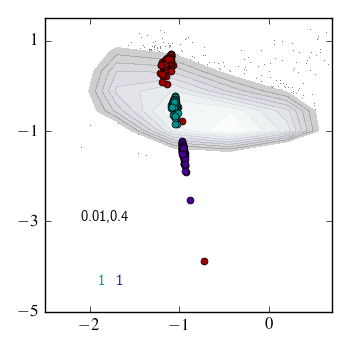}
     }
     \hspace*{-0.4cm}
     \subfloat{
       \includegraphics[width=0.252\textwidth]{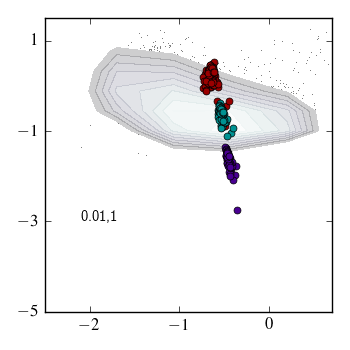}
     }
     \hfill
         \vspace*{-0.5cm}
          \subfloat{
       \includegraphics[width=0.252\textwidth]{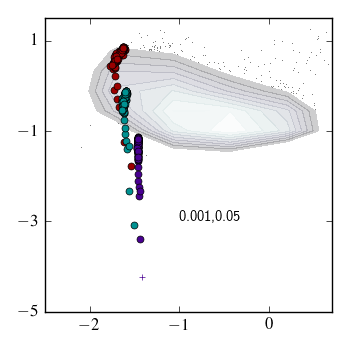}
     }
     \hspace*{-0.4cm}
     \subfloat{
       \includegraphics[width=0.252\textwidth]{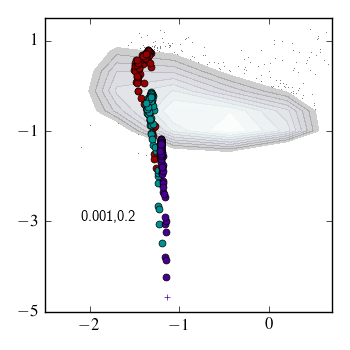}
     }
     \hspace*{-0.4cm}
          \subfloat{
       \includegraphics[width=0.252\textwidth]{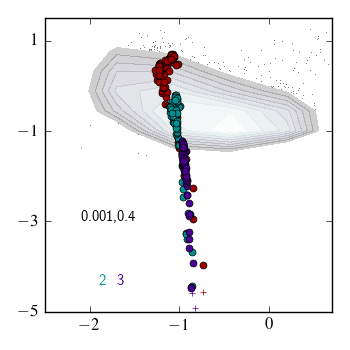}
     }
     \hspace*{-0.4cm}
     \subfloat{
       \includegraphics[width=0.252\textwidth]{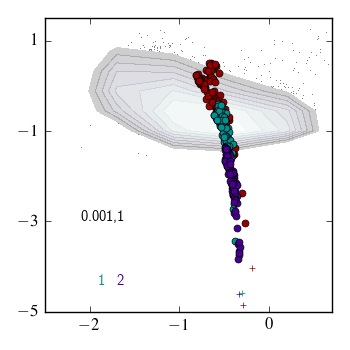}
     }
     \hfill
      \vspace*{-0.5cm}
          \subfloat{
       \includegraphics[width=0.252\textwidth]{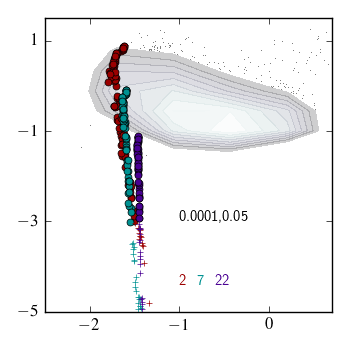}
     }
     \hspace*{-0.4cm}
     \subfloat{
       \includegraphics[width=0.252\textwidth]{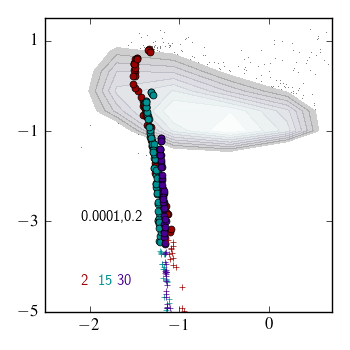}
     }
     \hspace*{-0.4cm}
          \subfloat{
       \includegraphics[width=0.252\textwidth]{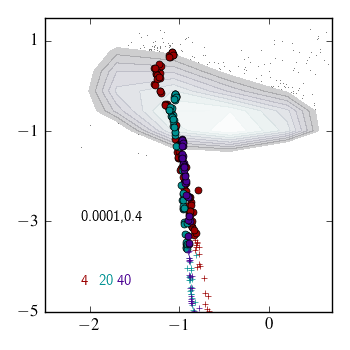}
     }
     \hspace*{-0.4cm}
     \subfloat{
       \includegraphics[width=0.252\textwidth]{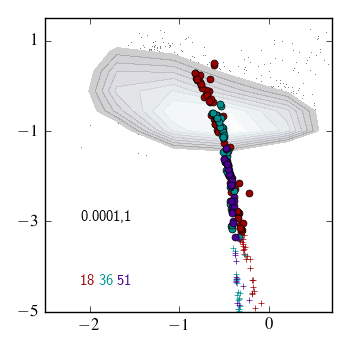}
     }
  \begin{minipage}{1.0cm}
  \vspace{0.5cm}
  \hspace{-11.1cm}
  \rotatebox{0}{ \large{log (\nii$\lambda$6584/\oii$\lambda$3727)}}
 \vspace{-0.5cm} 
 \hspace{11.1cm}
  \end{minipage}
     \vspace{0.5cm}
     \caption{Similar to Figure \ref{fig:lineratio1} but this time showing the \oiii/\oii\ versus  \nii/\oii\ diagram. Here again, we see the clear increase in the scatter in line ratios towards lower SFRs.}
     \label{fig:lineratio4}
   \end{figure*}

Figure \ref{fig:lineratio4} shows  \oiii{\ensuremath{\lambda}5007}/\oii{\ensuremath{\lambda}3727} versus \nii{\ensuremath{\lambda}6584}/\oii{\ensuremath{\lambda}3727}. Since the transition energy from N$^0$ to N$^+$ is similar to that required to ionise O$^0$ to O$^+$,  \nii{\ensuremath{\lambda}6584}/ \oii{\ensuremath{\lambda}3727} is mostly dependent on metallicity as highlighted by \citet{2002ApJS..142...35K}. Therefore, this predominant dependence on metallicity can be seen clearly in Figure \ref{fig:lineratio4}. The y-axis now shows a more metallicity independent measure of the ionisation energy of the stars. The $\log U=-2$ and the $\log U=-3$ models fall in the region where most SDSS galaxies lie, while the $\log U=-4$ values predict line ratios somewhat lower and higher than the typical SDSS galaxy shows. 

We interpret the increase of scatter in Figure \ref{fig:lineratio1} and \ref{fig:lineratio4} as a consequence of the stochastic sampling in the models, that results in more variation in the distribution of massive stars towards lower \SFR\ models. Below $\log$ \oiii/\hb\ $\approx$ -4 and $\log$ \oiii/\oii\ $\approx$ -4, the \oiii\ line flux is below our fiducial detection limit. This is predominantly the case for the \SFR\ = 0.0001 \Msun/yr models, but also for the models with the lowest values of \oiii/\hb\ and \oiii/\oii\ in the \SFR\ = 0.001 \Msun/yr bins. In the next sections we will therefore mainly focus on bins of models with this latter \SFR\ bin, because the line fluxes of these models reach the detection threshold and therefore the maximum detectable influence of calibrated metallicities.

\newpage 

\subsection{Stellar spectra and line ratios}
\begin{figure}
\captionsetup[subfigure]{labelformat=empty}
\centering
\subfloat[]{
  \includegraphics[width=85mm]{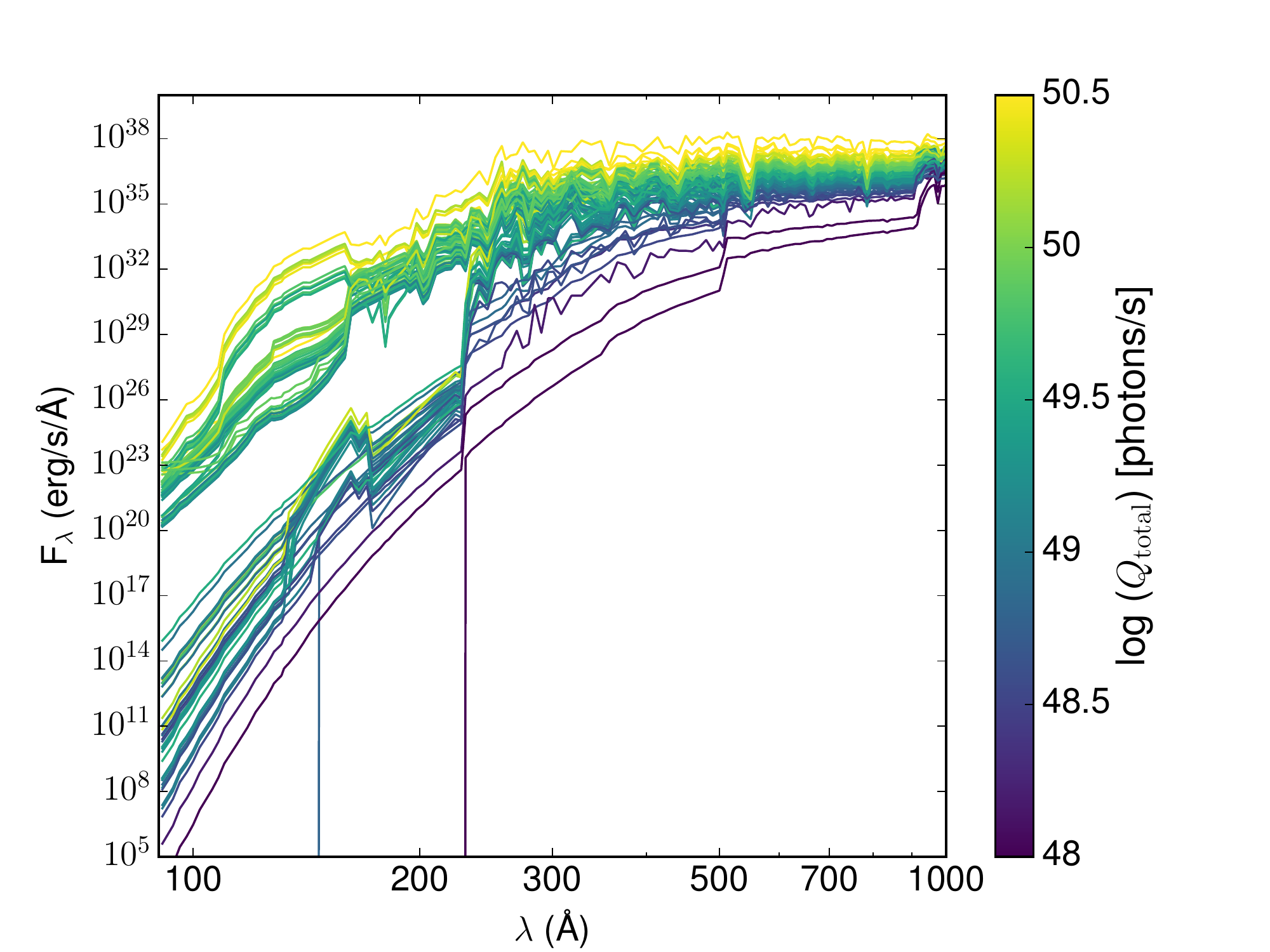}
}
\vspace{-0.92cm}
\hspace{0mm}
\subfloat[]{
  \includegraphics[width=85mm]{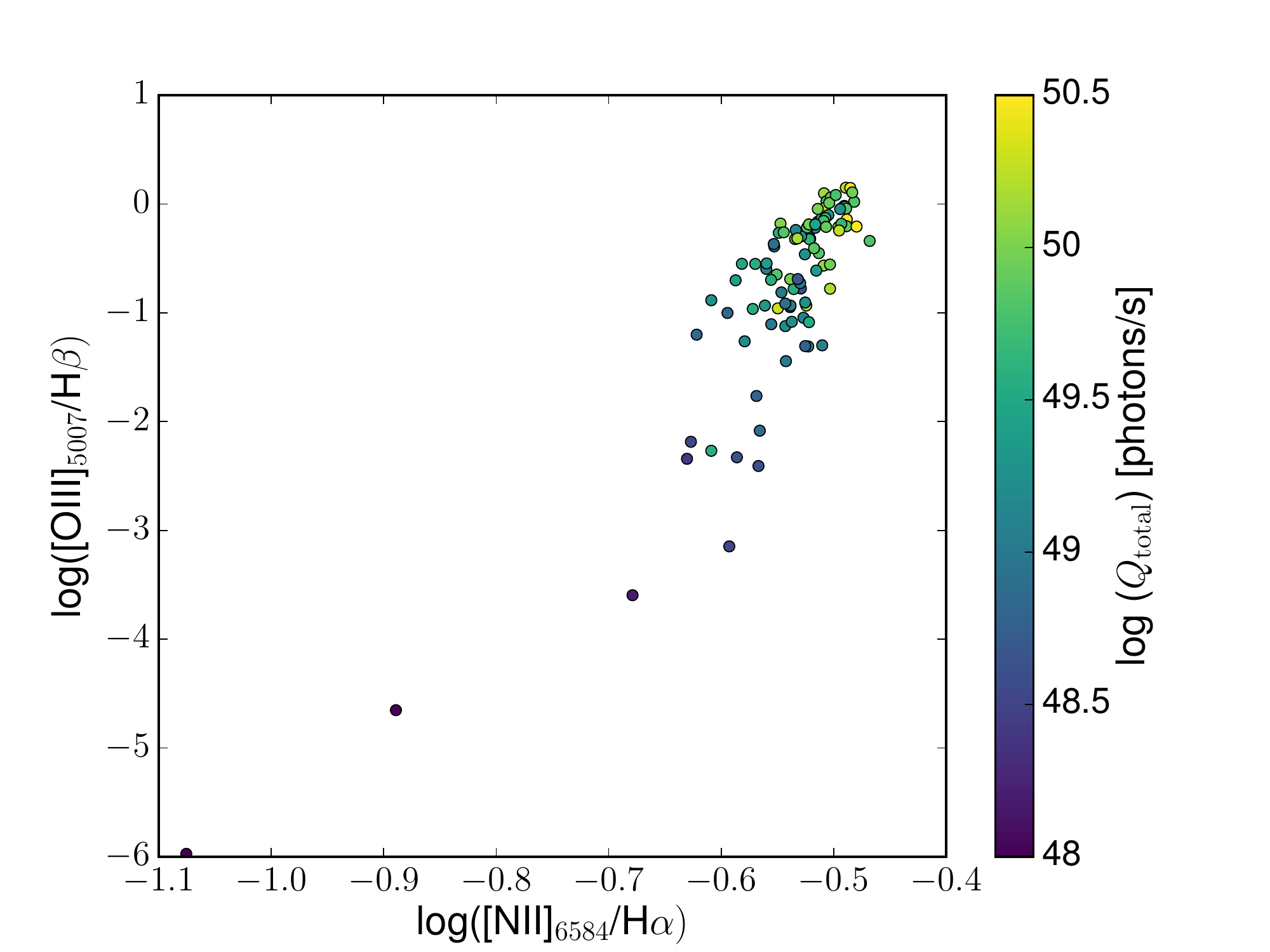}
}
\vspace{-0.93cm}
\hspace{0mm}
\subfloat[]{
  \includegraphics[width=85mm]{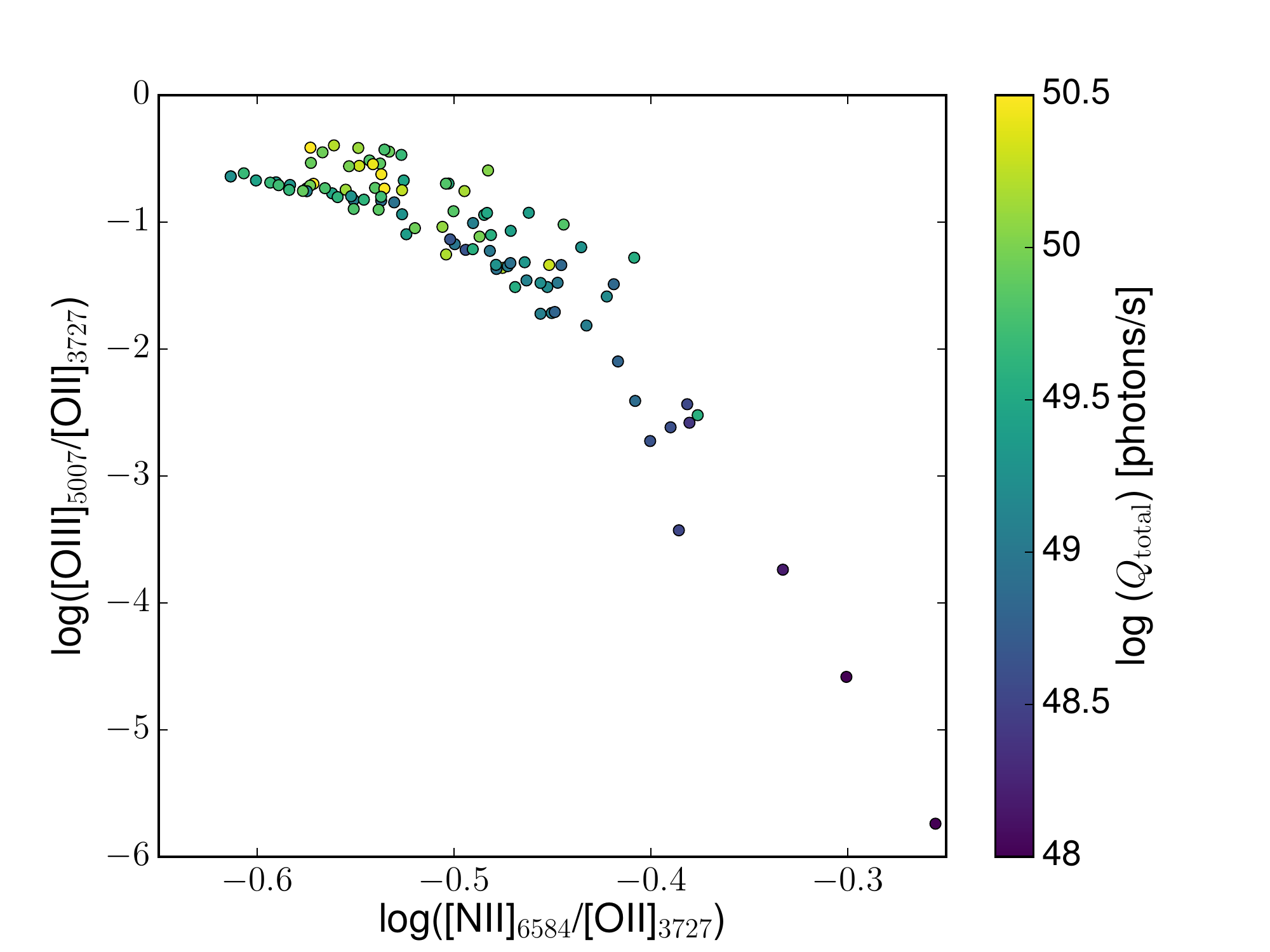}
}
\vspace{-0.7cm}
\caption{Results for our \Z\ = \Zsun, \SFR\ = 0.001 \Msun/yr and \logU\ = -3 models. Top panel: SED from our SLUG galaxy models, given in the range of energies higher than the Lyman limit ($<$912 $\AA$). The colour scheme is based on the total amount of ionising hydrogen photons emitted by these galaxies. Middle and lower panel: The \oiii{\ensuremath{\lambda}5007}/\hb\ versus \nii{\ensuremath{\lambda}6584}/\ha\ and the \oiii{\ensuremath{\lambda}5007}/\oii{\ensuremath{\lambda}3727}  versus  \nii{\ensuremath{\lambda}6584}/\oii{\ensuremath{\lambda}3727} diagnostic diagram for the same galaxies.}
\label{fig:specandbpt}
\end{figure}

In order to study the origin of the scatter in line ratios, as observed for low \SFR\ models in Figures \ref{fig:lineratio1} and \ref{fig:lineratio4}, we compare the stellar spectra from SLUG to the relative emission line ratios. In Figure \ref{fig:specandbpt} we show the spectra in combination with \oiii{\ensuremath{\lambda}5007}/\hb\ versus \nii{\ensuremath{\lambda}6584}/\ha\ and \oiii{\ensuremath{\lambda}5007}/\oii{\ensuremath{\lambda}3727}\ versus \nii{\ensuremath{\lambda}6584}/\oii{\ensuremath{\lambda}3727} diagrams of our \Z\ = 1 \Zsun, \SFR\ = 0.001 \Msun/yr and \logU\ = -3 sample. The spectra and line-ratio points are colour-coded to the total amount of ionising photons, \Q (E>13.6eV). In the top panel we show the spectrum bluewards of 1 Ry. The \oiii{\ensuremath{\lambda}5007}/\hb\ and the \nii{\ensuremath{\lambda}6584}/\ha\  line ratios decrease with a decreasing number of ionising photons (middle panel). We find a similar relation between ionising photon number and \oiii{\ensuremath{\lambda}5007}/\oii{\ensuremath{\lambda}3727}\ line ratio. However, the \nii{\ensuremath{\lambda}6584}/\oii{\ensuremath{\lambda}3727}\ ratio slightly decreases for a higher photon number, because the energy for the N$^0$ to N$^+$ transition is higher than the energy needed to singly ionise oxygen (14.5 and 13.6 eV respectively).

These figures show explicitly how transitions that have a higher energy requirement are progressively more affected by stochasticity. Since such line ratios are used in various metallicitiy calibrations, this variation will necessarily turn into a scatter in inferred metallicity at fixed true metallicity, a topic we now turn to. 


\section{Stochastic effect on metallicity determinations}
In this section we will study the scatter induced into gas-phase abundance estimates by stochastic IMF sampling. We will start by introducing four of the most common metallicity calibration methods. After that we will investigate the effect of a stochastic IMF on the result of each calibration. We test how this effect is related to the \SFR ,  the input metallicity, and the ionisation parameter.

\subsection{Metallicity calibrations}
We will here focus on two classes of metallicity estimators: the so-called \Te\ method which relies on auroral transitions in forbidden lines (in our case of oxygen) to estimate the electron temperature in the gas, and so-called strong-line methods (see  \citealt{2008ApJ...681.1183K} for an overview). The strong-line methods include both theoretical calibrations (e.g. \citealt{1991ApJ...380..140M,2002ApJS..142...35K} for an overview), as well as empirically calibrated methods (e.g.  \citealt{2001A&A...374..412P, 2004MNRAS.348L..59P}). There is substantial variation between different metallicity indicators, for instance, theoretical estimates usually predict up to 0.6 dex higher metallicities compared to temperature-sensitive empirical predictions \citep{2006ApJ...652..257L}.

As is well known, and explicitly demonstrated above, emission line ratios are sensitive not only to the metal content of the ionising gas, but also to the ionisation parameter. However the calibrations used in the literature implicitly or explicitly assume a fully sampled IMF and we will explore the consequences of this below. 

While the model described above provides a reasonable description of SDSS galaxies, it is not the result of a rigorous optimisation and uses different assumptions/software from the models used for strong-line calibrations. We are therefore likely to see small systematic offsets between the metallicities predicted by the strong-line methods and the input metallicities in our models. Our focus is however on the \textit{relative} effects caused by a stochastically sampled IMF and we will therefore focus on the scatter and relative trends of the derived metallicity relative to that of a non-stochastic model, e.g. a model with a fully sampled IMF, with identical properties.
\subsubsection{Direct \Te\ measurements}
\label{s_te}
After collisional excitation, the downward transition for the electron back to the ground state produces photons that can give rise to 'forbidden' emission lines. The intensity of these lines is dependent on the electron temperature \Te , the density of the gas and the chemical composition. Therefore, knowing the temperature and the density of the gas gives insight into the metallicity. The \Te\ of the \oiii\ region is related by the the difference between the strengths of the strong temperature dependent auroral line \oiii{\ensuremath{\lambda}4363} and the \oiii{\ensuremath{\lambda\lambda}4959,5007} lines, by approximately:
\begin{equation}
\frac{j_{\lambda 4959}+j_{\lambda 5007}}{j_{\lambda 4363}} = \frac{7.9 \mathrm{exp} (3.29 \times 10^4 /\Te)}{1+4.5 \times 10^{-4} n_e /\Te ^{1/2}}
\end{equation}
where $j_{\lambda}$ are the emission line luminosities, see \citet{1989agna.book.....O}. We computed \Te(\oiii) using the ${\tt nebular.ionic}$ routine in Pyneb \citep{2015A&A...573A..42L}. 

Although the auroral \oii{\ensuremath{\lambda}7325} line has been detected at high metallicity regions, either directly \citep{2016ApJ...827..126B, 2015ApJ...806...16B} or in stacked spectra \citep{2017MNRAS.465.1384C}, we adopted the relation beween \Te(\oiii) and \Te(\oii) from \citet{2006A&A...448..955I}  to calculate the electron temperature in the \oii\ region. The reason for this is that the \oii{\ensuremath{\lambda}7325} line is usually weak and therefore challenging to measure in observed spectra. Moreover, since the difference  between the emitted wavelengths of \oii{\ensuremath{\lambda}3727} and \oii{\ensuremath{\lambda}7325} is relatively large, accurate dust measurements are necessary to correct the \oii\ lines for interstellar dust absorption. 

We used Pyneb to calculate the the O$^{+}$ and O$^{++}$ abundances using \Te(\oii) and \Te(\oiii) in combination with a constant electron density  (n$_e$ = 100 cm$^{-3}$) . Armed with this we can then write the total oxygen abundance as
\begin{equation}
\frac{\mathrm{O}}{\mathrm{H}} = \frac{\mathrm{O}^+}{\mathrm{H}^+} + \frac{\mathrm{O}^{++}}{\mathrm{H}^{++}}.
\end{equation}
This neglects the contribution of O$^{3+}$, that can be found in highly ionised gas, but this is minimal for galaxies with ionisation energies such as those in our models \citep{2013ApJ...765..140A}. 

At higher metallicities, the \oiii\ line ratio becomes very challenging to measure due to the increasing weakness of the \ensuremath{\lambda}4363 line. This then is the regime of the strong-line metallicity calibrations to which we turn next.
\subsubsection{\ntot \space metallicity calibrations}
\label{s_n2o2}
The ratio between \nii \space and \oii \space is an example of a metallicity estimator that is less affected by a difference in the ionising energy distribution, because the ions have similar ionisation energies. It is  strongly dependent on metallicity for the reason that, in case of \Z\ > 0.5 \Zsun,  \nii \space is predominantly a secondary element, and the flux therefore scales more strongly with metallicity than the \oii \space line \citep{1979A&A....78..200A, 2000A&A...356...89C}. Note that the \ntot\ method relies on a tight N/O relationship, and while most galaxies follow this, it is well established that a subset of galaxies deviate from this \citep{2017arXiv170407604C,2017arXiv170303813B,2015MNRAS.449..867B,2014MNRAS.444..744Z}. Here we will make the explicit assumption that galaxies fall on such a tight N/O relationship and ignore this complication, but this will therefore overstate somewhat the power of the \ntot\ method on a sample of real galaxies.

Below 0.5 \Zsun\ the \nii/\oii \space metallicity calibration is not useful, because the metallicity dependence is lost since nitrogen is predominantly a primary nucleosynthesis element in this metallicity range. For this paper, we use the theoretically determined \ntot\ calibration from \citet{2002ApJS..142...35K}, given by:
\begin{equation}
   \begin{aligned}
     \mathrm{log(O/H)} +12 = {} & \mathrm{log} [1.54020 +  1.26602 \times \mathrm{\ntot} \\
	     & + 0.167977 \times \mathrm{\ntot}^{2}] + 8.93 
   \end{aligned}
\end{equation}
where \ntot\ = log (\nii/\oii). We will adopt this relation also for models with low initial gas-abundances, but caution that it is less powerful as a metallicity indicator in this regime for the reasons outlined above. In practice, the \ntot\ method is of limited use at high redshift because it requires spectra covering a long range in wavelength, including also at least two Balmer lines to accurately correct the \oii{\ensuremath{\lambda}3727}\ and \nii{\ensuremath{\lambda}6584}\ lines for internal dust reddening.

\subsubsection{\rtt \space metallicity calibrations}
\label{s_r23}
A method for metallicity determinations using  both \oii\ and \oiii\ lines was formulated in the studies of \citet{1979A&A....78..200A, 1979MNRAS.189...95P, 1980MNRAS.193..219P}. The \rtt\ line-ratio is defined by 
\begin{equation}
\mathrm{\rtt} \space = \space \mathrm{log} (\frac{ \oii{\ensuremath{\lambda}3727} +  \oiii{\ensuremath{\lambda}4959} +  \oiii{\ensuremath{\lambda}5007}}{\hb}).
\end{equation} 
Numerous studies have been performed on the use of this line ratio for metallicity determinations, both from the empirical \citep{1979MNRAS.189...95P, 1980MNRAS.193..219P, 2001A&A...374..412P, 2005ApJ...631..231P} and theoretical \citep{1991ApJ...380..140M, 1994ApJ...420...87Z, 2002ApJS..142...35K, 2004ApJ...617..240K} points of view. 
However, in addition to being sensitive to metallicity, the \rtt\ ratio is also sensitive to the ionisation energy of the source, for which an estimate can be obtained from the ratio of the  \oii\ to \oiii\ lines, \ott , as given by
\begin{equation}
\mathrm{\ott} = \mathrm{log} (\frac{\oiii{\ensuremath{\lambda}4959} + \oiii{\ensuremath{\lambda}5007}}{\oii{\ensuremath{\lambda}3727}}).
\end{equation}
One disadvantage of this method is that \rtt\ viewed as a function of metallicity is double-valued. With the upper branch corresponding to the high metallicity solution and the lower branch corresponding to the low metallicity solution. This method therefore requires additional emission line measurements in order to break the degeneracy between the upper and lower branches. In this paper we use the calibration published by \citet{1991ApJ...380..140M}, because it is arguably the most well-studied of the \rtt\ methods. This is given by
\begin{equation}
\begin{aligned}
\mathrm{12+log(O/H)}_{\mathrm{upper}} = {} & 9.061 - 0.2 \mathrm{\rtt} - 0.237 \mathrm{\rtt}^{2} \\
    & - 0.305 \mathrm{\rtt}^{3} - 0.0283 \mathrm{\rtt}^{4} \\
    & - \mathrm{\ott}(0.0047-0.0221 \mathrm{\rtt} - 0.102 \mathrm{\rtt}^{2} \\
    & - 0.0817 \mathrm{\rtt}^{3} - 0.00717 \mathrm{\rtt}^{4})
\end{aligned}
\end{equation}
\begin{equation}
\begin{aligned}
\mathrm{12+log(O/H)}_{\mathrm{lower}} = {} & 7.056 + 0.767 \mathrm{\rtt} \\
  & - \mathrm{\ott}(0.29+0.332 \mathrm{\rtt} - 0.331 \mathrm{\rtt}^{2}).
\end{aligned}
\end{equation}
While observationally it can be challenging to break the degeneracy, in our case we know which branch we should use so we will not concern ourselves with this challenge here. We apply the upper branch our models with \Z\ = 1.0 \Zsun\ and the lower branch for our models with \Z\ = 0.005, 0.2 and 0.4 \Zsun. 
\subsubsection{\otnt \space metallicity calibrations}
\label{s_o3n2}
The last line ratio calibration that we will test on our models is \oiii/\nii, referred to as \otnt, initially brought up as an estimation of the metallicity by \citep{1979A&A....78..200A}. Later, this line ratio metallicity determination was improved using a larger empirical library by  \citet{2004MNRAS.348L..59P} and is defined by
\begin{equation}
\mathrm{\otnt} = \mathrm{log}\frac{\oiii{\ensuremath{\lambda}5007}/\hb}{\nii{\ensuremath{\lambda}6584}/\ha}.
\end{equation}   
The ratio \ha/\hb\ is added to minimise the effect of reddening by dust from the interstellar matter. This line ratio is only sensitive to metallicity in the range -1 < \otnt < 1.9, which corresponds to a metallicity range of 8.12 < 12+log(O/H) < 9.05. The metallicity relates to \otnt\ through \citep{2004MNRAS.348L..59P}
\begin{equation}
\mathrm{12+log(O/H)} = 8.73 - 0.32 \times \mathrm{\otnt}.
\end{equation}
Since the difference between ionisation energy of \nii \space and \oiii \space is large, this estimator is dependent on the energy distribution in the ionising part of the spectral energy distribution of the central source and thereby also to variation in the energy distribution. Variations in the relative number of massive stars is potentially a problem for the determined metallicity, which we shall demonstrate in the next section. 

\subsection{Influence of SFR}
\begin{figure*}
\captionsetup[subfigure]{labelformat=empty}
\centering
\subfloat[]{
  \includegraphics[width=93mm]{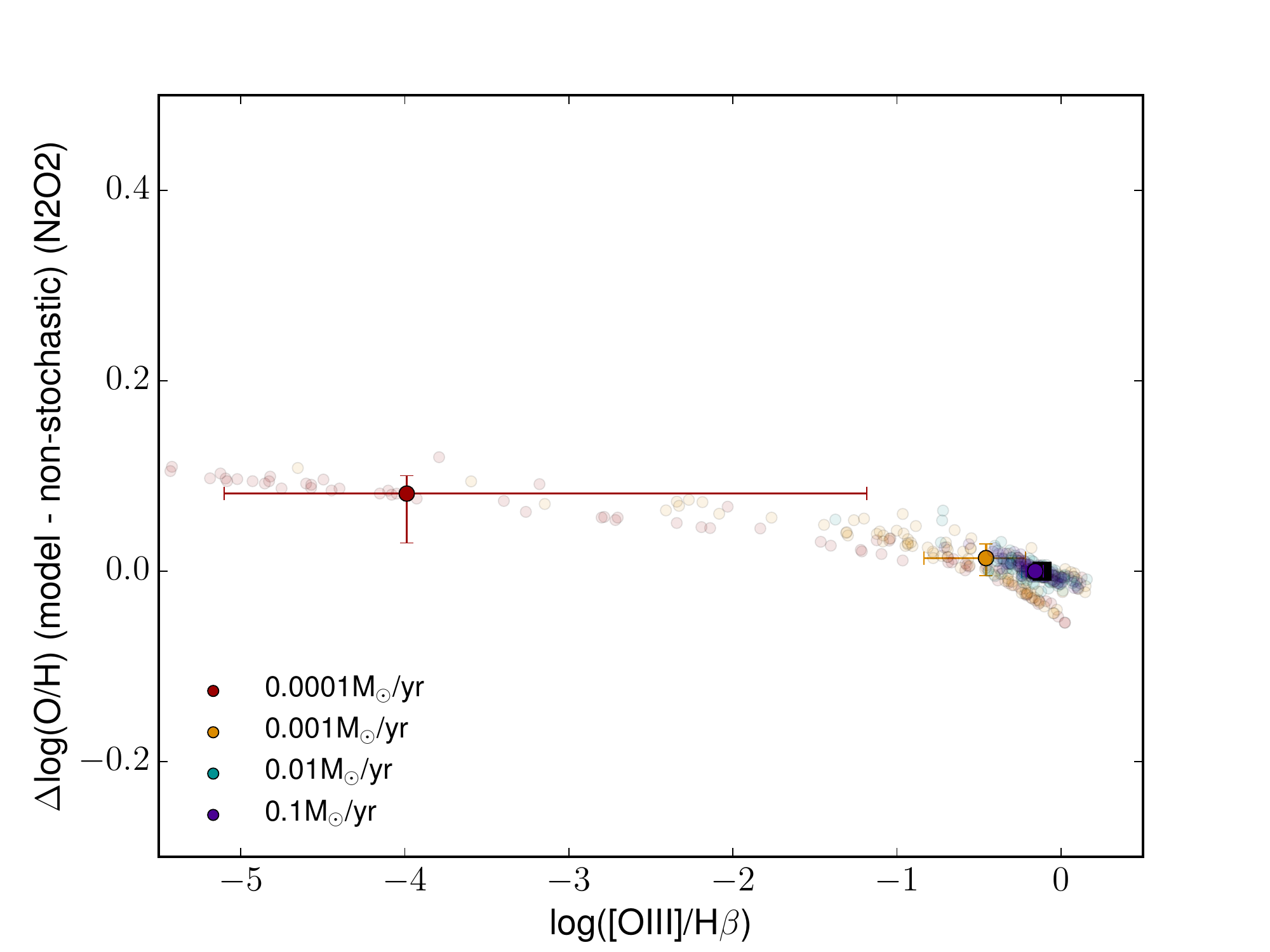}
}
\subfloat[]{
  \includegraphics[width=93mm]{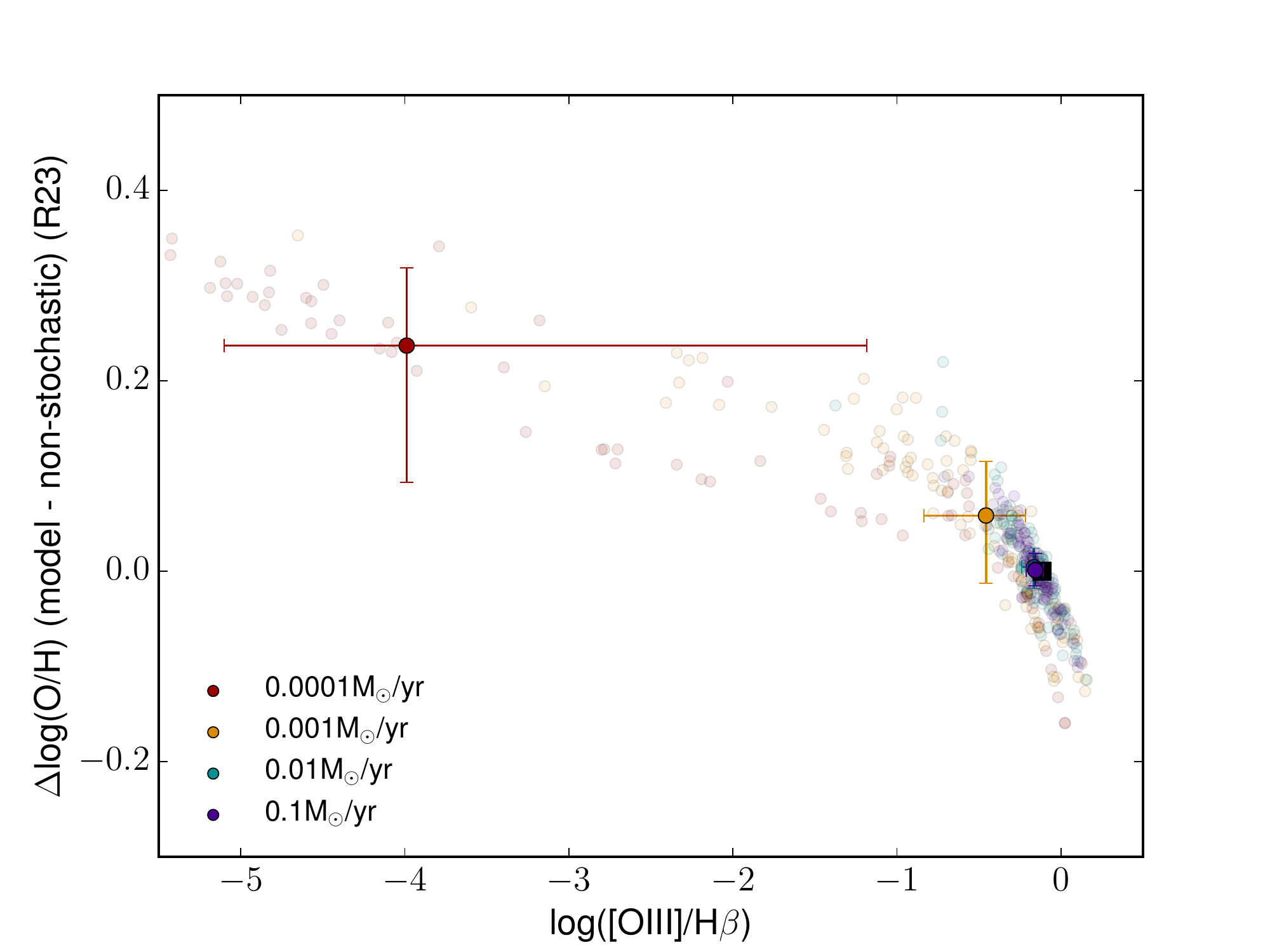}
}
\hspace{0mm}

\subfloat[]{
  \includegraphics[width=93mm]{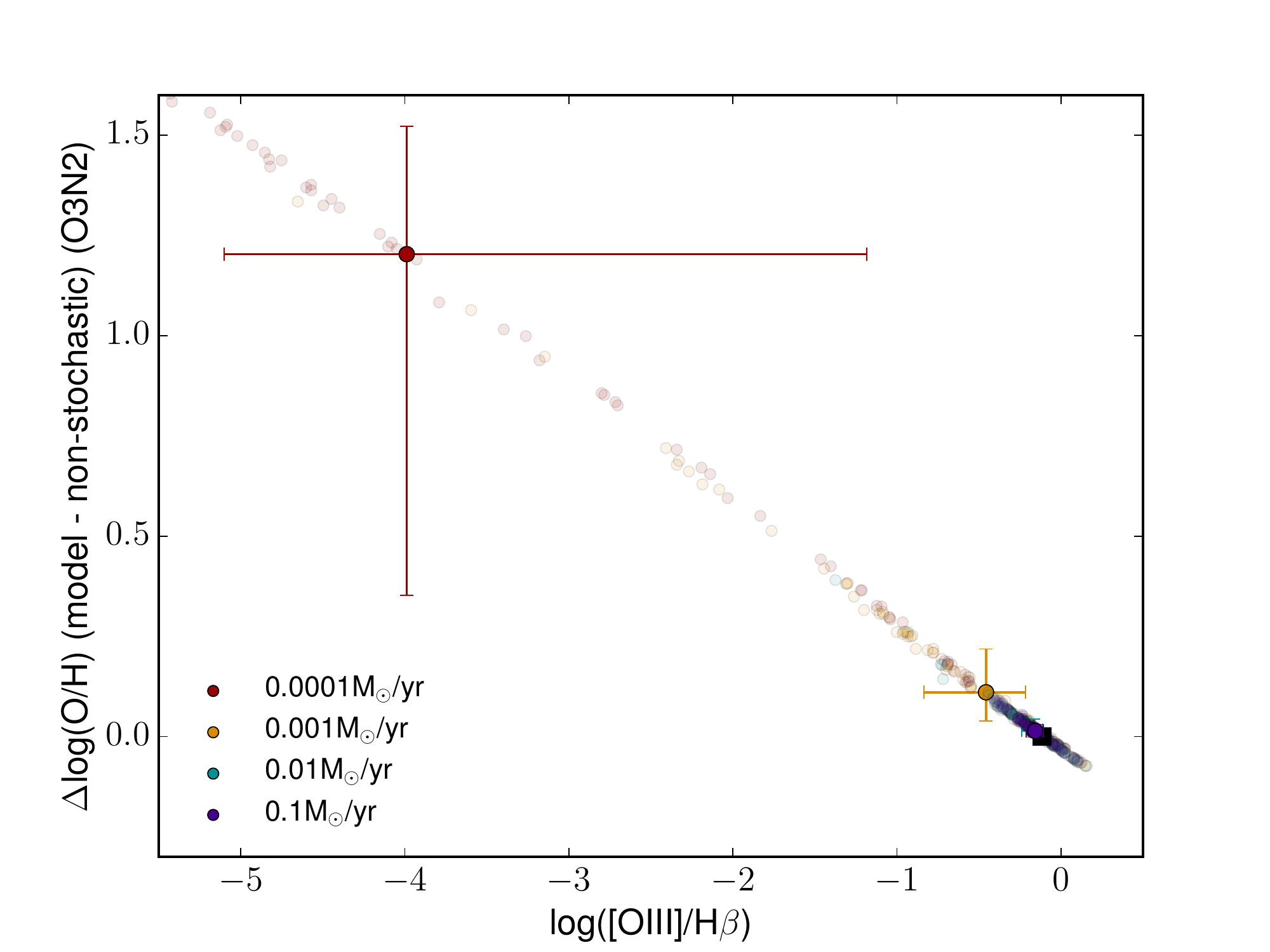}
}
\subfloat[]{
  \includegraphics[width=93mm]{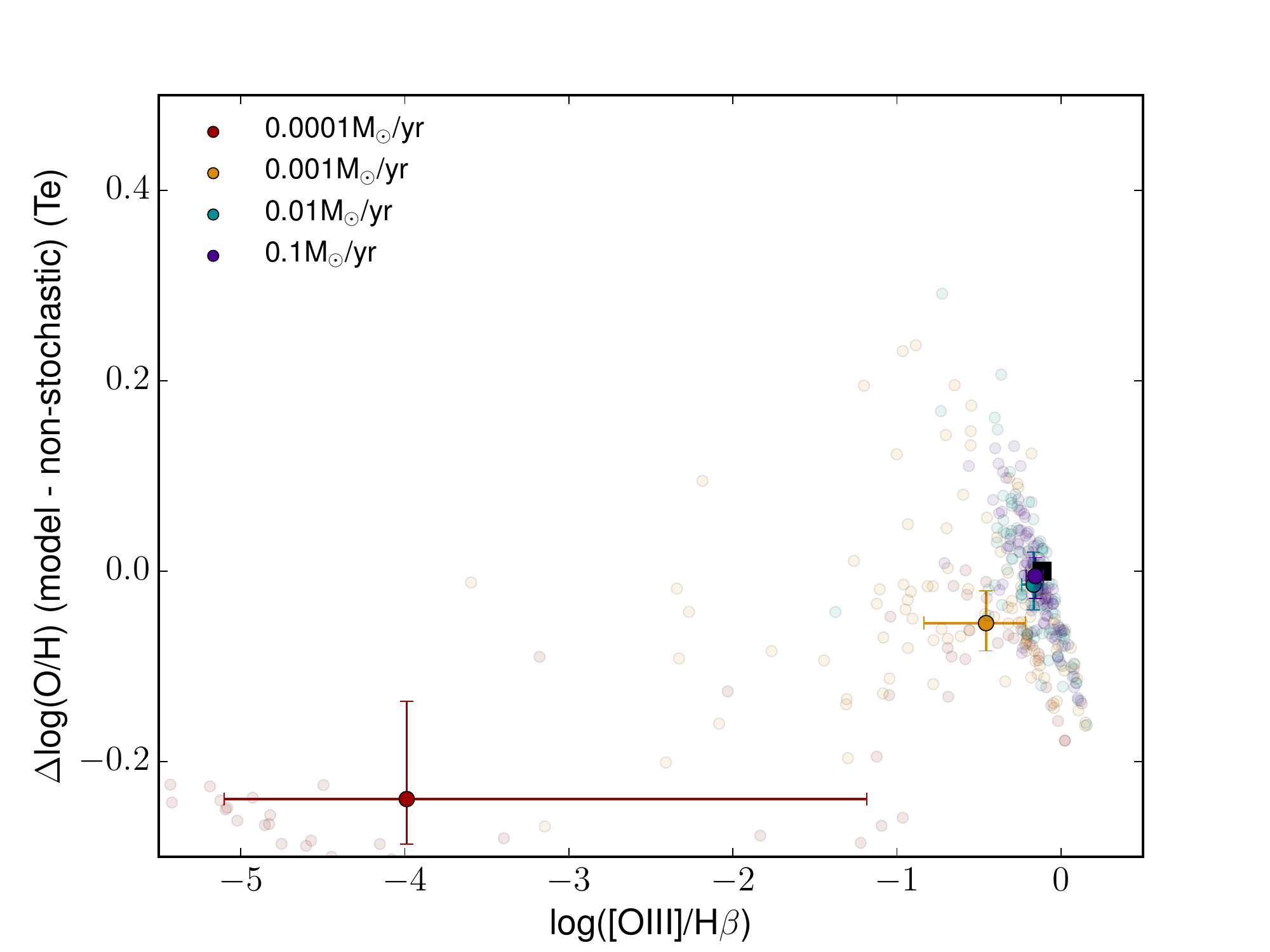}
}
\caption{log(O/H) of the stochastic models relative to the log(O/H) of a typical non-stochastic model, versus the line ratio  \oiii{\ensuremath{\lambda}5007}/\ha. The metallicities are derived with the \ntot\ (top left), the \rtt\ (top right) and the \otnt\ (bottom left) calibrators, and the direct \Te\ method (bottom right). The input \SFR\ of the models is 0.0001 (red), 0.001 (orange), 0.01 (cyan) and 0.1 (purple) \Msun/yr and for all models we adopted \logU\ = -3 and input \Z\ = 1 \Zsun. The black squares represent the results of the non-stochastic model, that is the same for all \SFR\ bins. The transparent coloured points show the results of the models, and the opaque points the median of a \SFR\ bin. The error bars show the distribution of 68 $\%$ of the models.}
\label{fig:resultsfr}
\end{figure*}
\begin{figure}
 \begin{minipage}{1.0cm}
  \vspace*{4cm}
  \hspace*{-4.2cm}
  \rotatebox{90}{ \large{$\Delta$log(O/H)}}
 \vspace*{-4cm} 
 \hspace*{4.2cm}
  \end{minipage}%
\vspace*{-2cm}  
\centering
 \subfloat{
       \includegraphics[width=0.9\columnwidth]{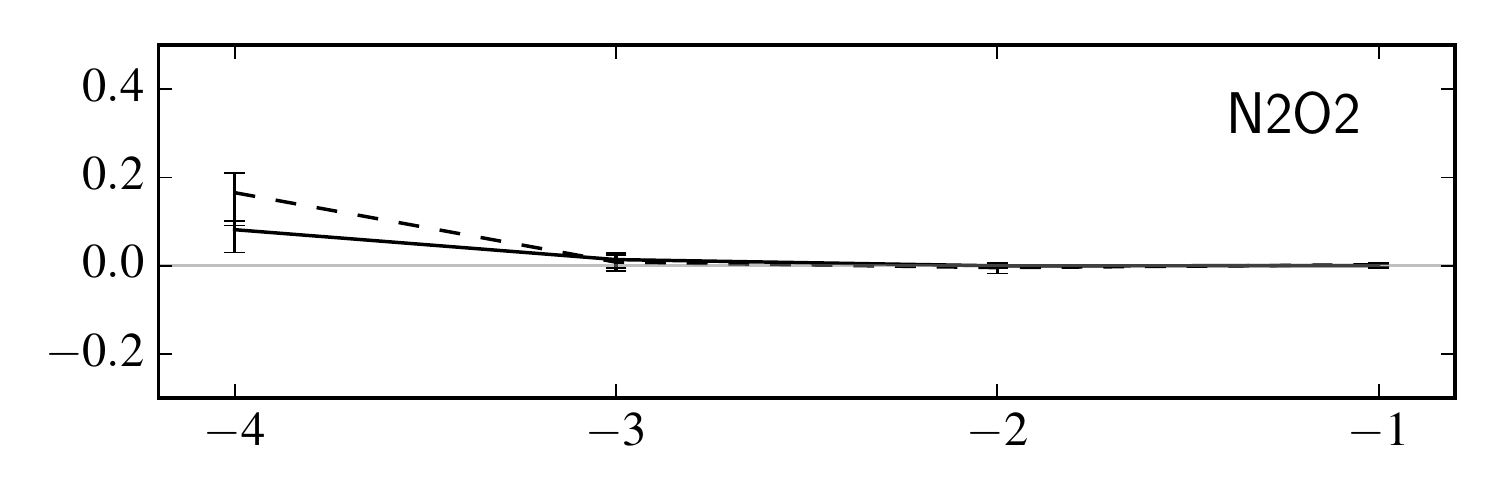}
     }
     \hfill
      \vspace*{-0.63cm}
     \centering
     \subfloat{
       \includegraphics[width=0.9\columnwidth]{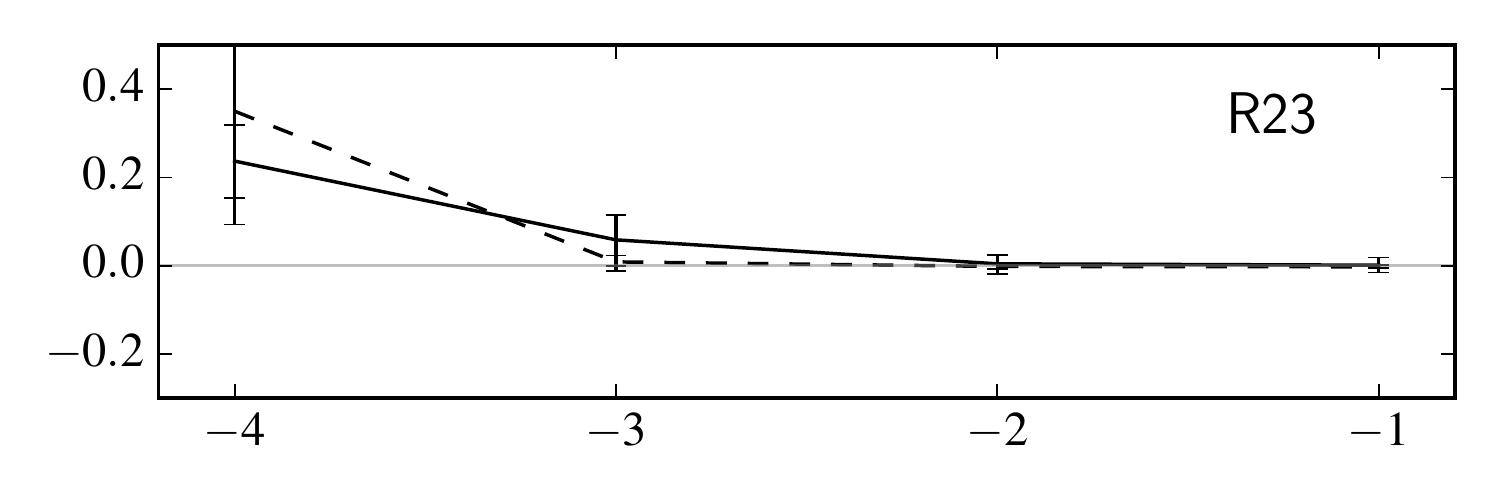}
     }
      \hfill
       \vspace*{-0.63cm}
      \centering
      \hspace*{0.08cm}
          \subfloat{
       \includegraphics[width=0.881\columnwidth]{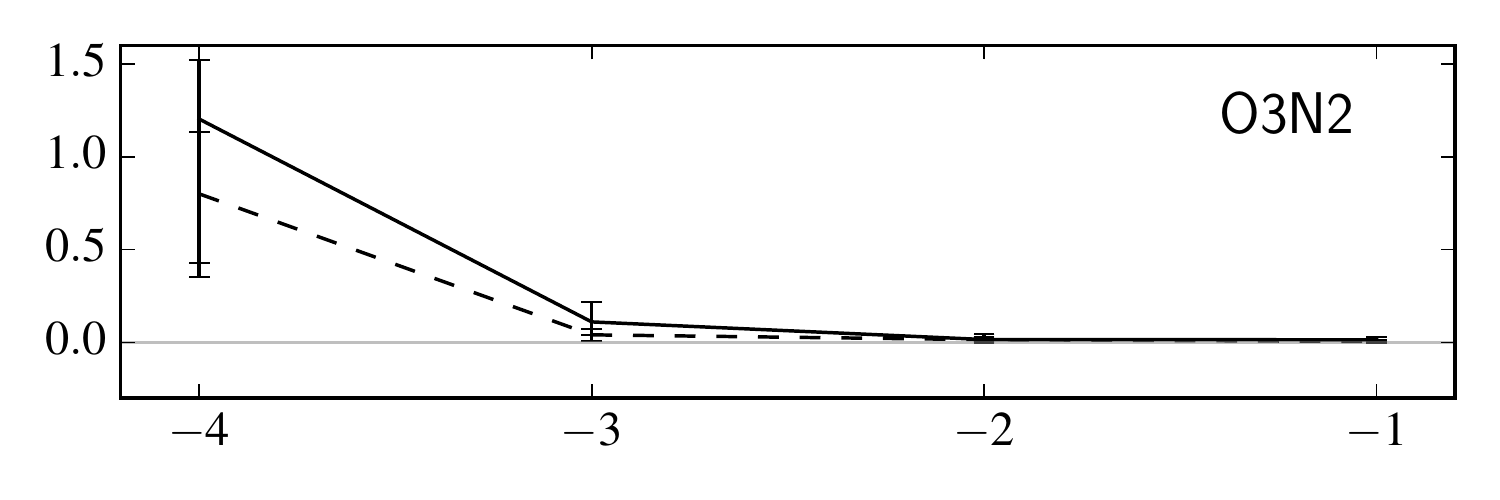}
     }
     \hspace*{-0.08cm}
     \hfill
      \vspace*{-0.63cm}
     \centering
     \subfloat{
       \includegraphics[width=0.9\columnwidth]{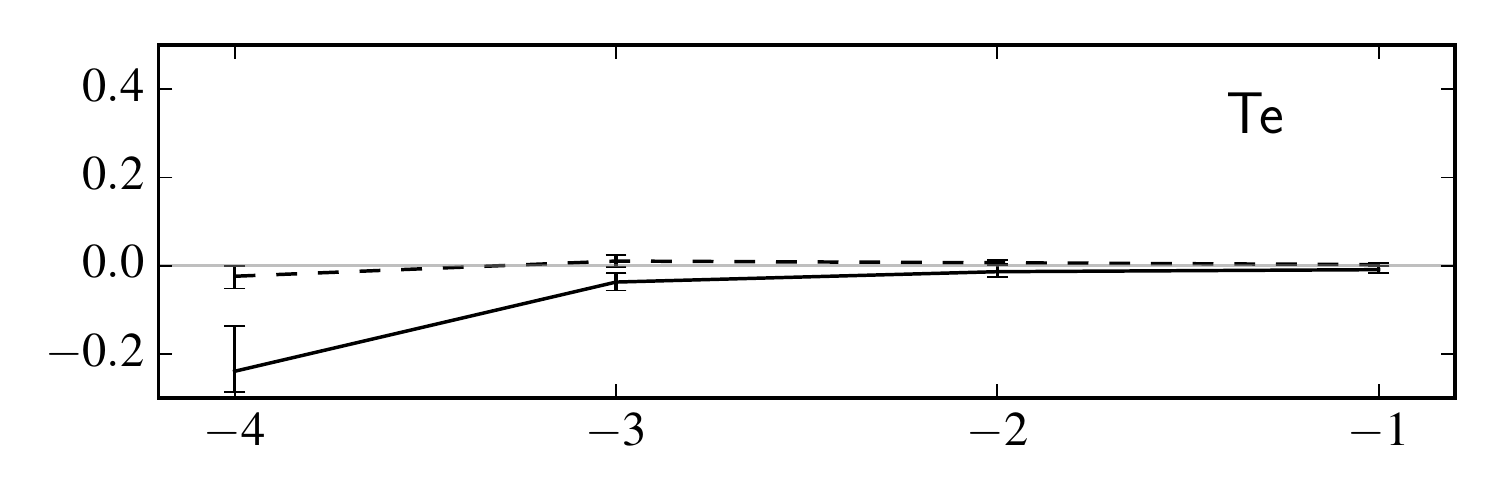}
     }
      \hfill
      \vspace*{-0.15cm}
   \begin{minipage}{1.0cm}
  
  \hspace*{-1.5cm}
  \rotatebox{0}{\large{log(Input SFR (\Msun /yr))}}
 
 \hspace*{0.8cm}
  \end{minipage}%
\caption{Summary of the relation between the relative metallicity and the input \SFR\ for metallicities derived by, from top to bottom, the \ntot , \rtt , \otnt\ and the \Te\ method. The solid black line shows the results of models with \Z\ = \Zsun , which is the same as the metallicity of the models in Figure \ref{fig:resultsfr}, and the dashed line if for models with \Z\ = 0.2 \Zsun. The error bars indicate the same 1-$\sigma$ spread in relative metallicity as in Figure \ref{fig:resultsfr}. Note that the y-axis range of the panel that shows the \otnt\ deviates from the rest.}
\label{fig:summarysfr}
\end{figure}
In the top left panel of Figure \ref{fig:resultsfr} we compare the oxygen abundance implied from the \oii/\nii\ ratio using the calibration discussed in Section \ref{s_n2o2}, for stochastic models, to the oxygen abundance of a non-stochastic model with the same properties on the y-axis, which we will refer to as the relative metallicity in the following. We plot this against the  \oiii{\ensuremath{\lambda}5007}/\hb\ that we saw above is rather sensitive to the stochastic effects in the IMF sampling. The total ionisation potential and the input metallicity are identical for all these models, with $\log U(total)$ = -3 and \Z\ = \Zsun . Each coloured, partially transparent point corresponds to the result of a single stochastic model, where the colour indicates the different input \SFR\ of the model (see legend). The solid dots show the mean for each SFR with the error bars indicating the 1-$\sigma$spread in each direction. The error bars are meant only for illustrative purposes and do not indicate uncertainties and in particular they are not independent. The solid squares indicate the results of a non-stochastic model in each \SFR\ bin. We show the relative metallicity using the \rtt\ calibration as described in Section \ref{s_r23} in the top right panel. The same results obtained from the \oiii/\nii\ ratio (Section \ref{s_o3n2}), and \Te\ (Section \ref{s_te}) are shown in the bottom left and bottom right panel respectively. We point out the the axis ranges are the same for all but the \otnt\ plot, for which the y-axis covers a larger range due to more scatter of the relative metallicity for this calibration. 

The median log(O/H) determined by the \ntot\ calibration agrees well with the metallicity of the non-stochastic model (only the median relative metallicity of the models with \SFR\ = 0.0001 \Msun/yr is 0.08 dex off) and the scatter is also small. Thus we conclude that this calibration is, as expected, only very weakly sensitive to stochastic effects. 

The other panels show the results for the other indicators. We see similar offsets (up to $\sim$ 0.2 dex for the lowest \SFR\ bin) in relative metallicities for the results obtained with the \rtt\ (top right) and the \Te\ method (bottom right), although the former is towards higher and the latter towards lower metallicities. We derived the largest spread in relative metallicities for abundances derived by the \otnt\ method, reaching an median offset of $\approx 1.2$ dex for the \SFR\ = 0.0001 \Msun/yr model. We will discuss the reason for this strong offset below. 

We summarise the relation between relative metallicity and \SFR\ for the four different methods in Figure \ref{fig:summarysfr} for our results of the \ntot , \rtt , \otnt\ and \Te\ calibrations from top to bottom panel (black solid line). We also added the same results for models with \Z\ = 0.2 \Zsun\ (dashed line), because metallicities calibrated by the \Te\ method are doubtful at solar metallicity. As in Figure \ref{fig:resultsfr}, the y-axis range of the \otnt\ panel deviates from the rest. It is notable that at modest star formation rates, $\log SFR \ge -2$, the effects of stochasticity can be neglected, but at lower SFR any indicator that includes [O III] shows both a bias and an increasing scatter.
\subsection{Influence of input metallicity}
\begin{figure*}
\captionsetup[subfigure]{labelformat=empty}
\centering
\subfloat[]{
  \includegraphics[width=93mm]{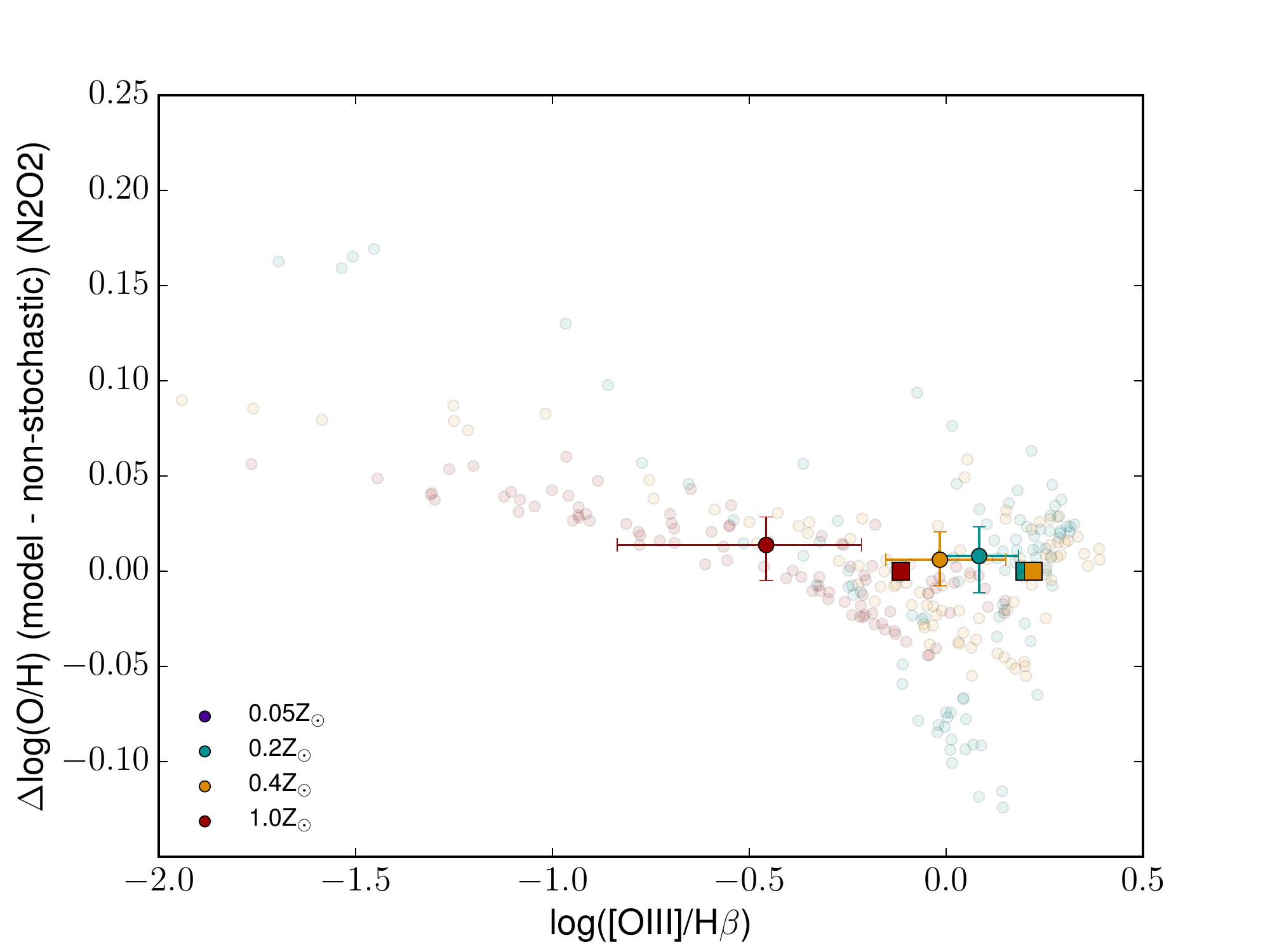}
}
\subfloat[]{
  \includegraphics[width=93mm]{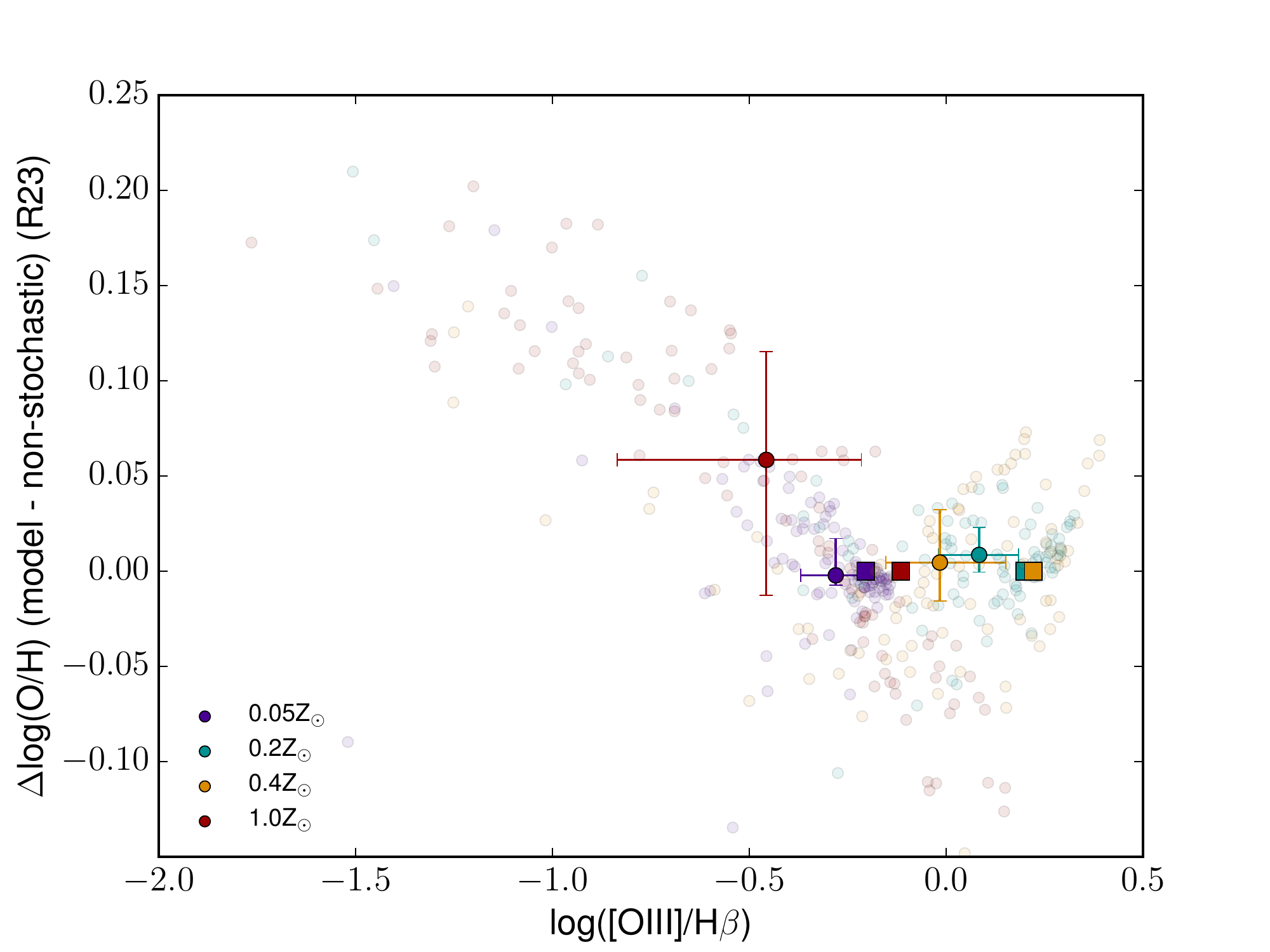}
}
\hspace{0mm}
\subfloat[]{
  \includegraphics[width=93mm]{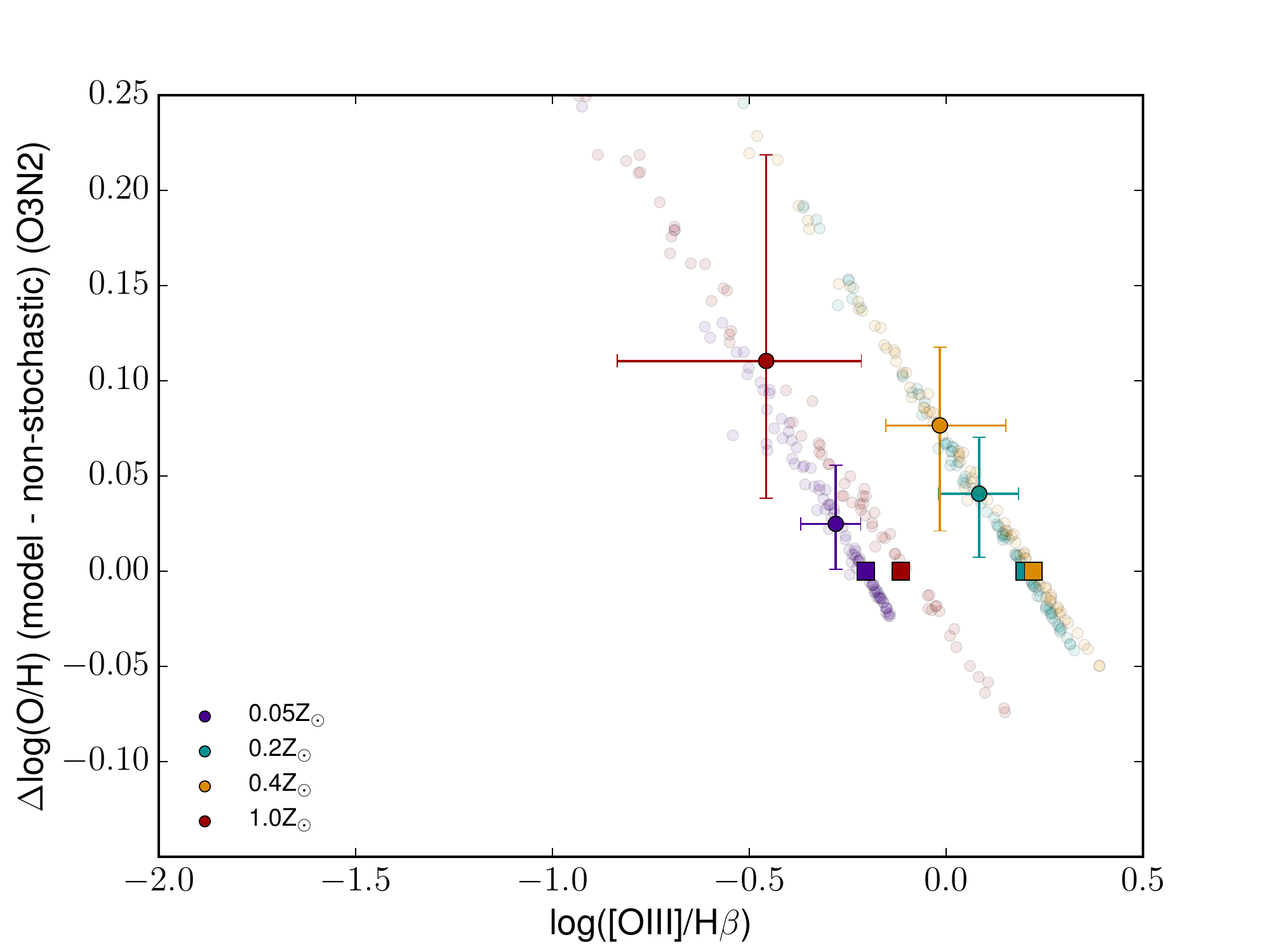}
}
\subfloat[]{
  \includegraphics[width=93mm]{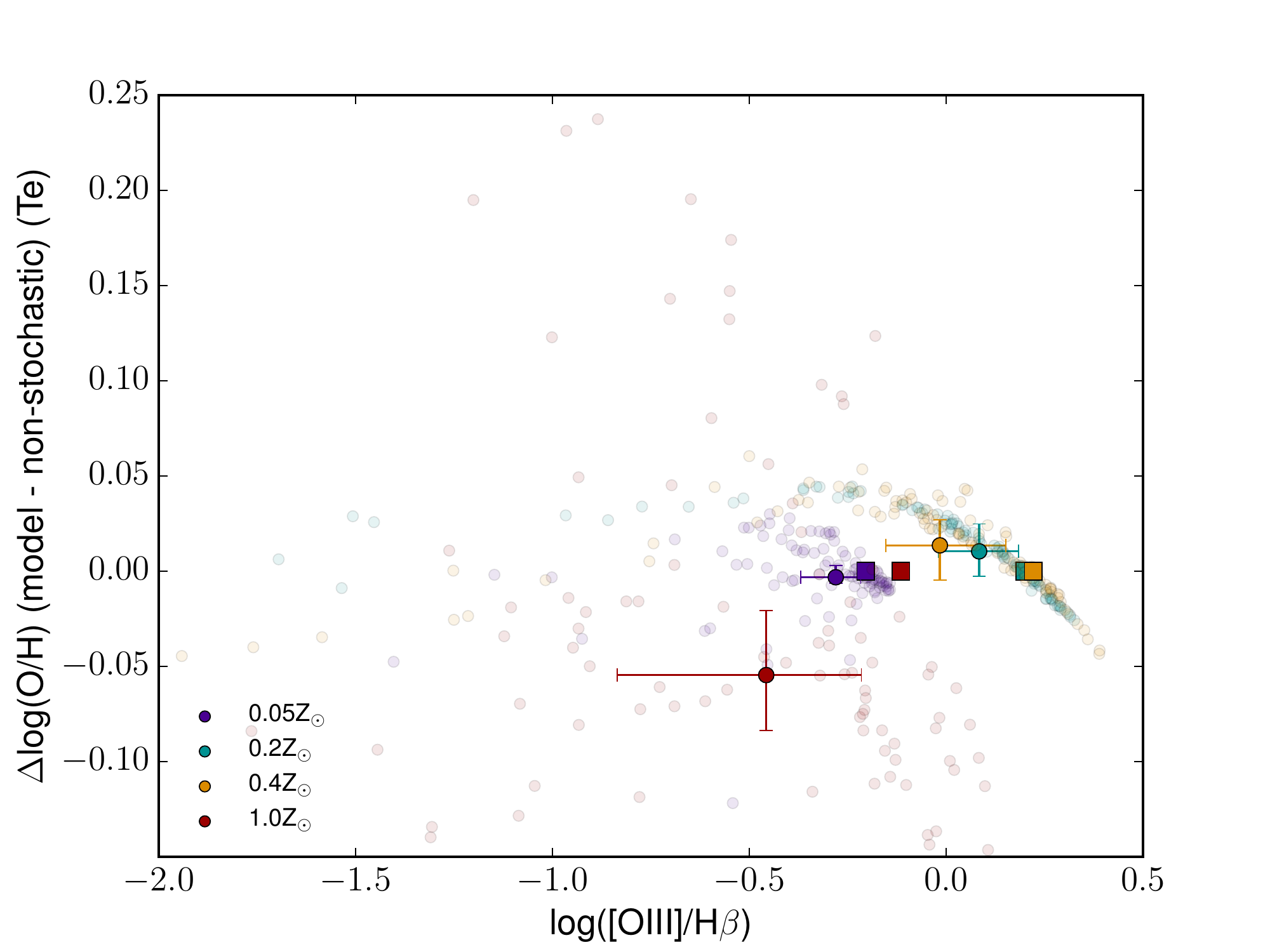}
}
\caption{Similar to Figure \ref{fig:resultsfr}, but the different colours match the input \Z, with \Z\ = 0.05 \Zsun (purple), \Z\ = 0.2 \Zsun (cyan), \Z\ = 0.4 \Zsun\ (orange) and \Z\ = 1 \Zsun (red). In contrary with the results in Figure \ref{fig:resultsfr}, the non-stochastic models are not the same for every \Z\ bin and presented by the coloured squares. The input \SFR\ of all models is 0.001 \Msun/yr and \logU\ = -3.}
\label{fig:resultmetal}
\end{figure*}
In this section we investigate how the input metallicity affects the calibrated abundances. In Figure \ref{fig:resultmetal} we show similar plots as in Figure \ref{fig:resultsfr}, but here we split the sample in input metallicity bins. The star formation rate and the ionisation parameter are constant over the models and equal \SFR\ = 0.001 \Msun/yr and \logU\ = -3. The results of the non-stochastic models are indicated by the filled, coloured squares. 

In the top left panel we show the results of the \ntot\ calibrator. Again, these derived metallicities are consistent with that of the non-stochastic models at all metallicities, with a maximum relative scatter of 0.04 dex and a tiny offset between the median and non-stochastic results. We make a similar statement about the \Te\ (bottom right) derived abundances, for which 68 $\%$ of the results are within 0.05 dex. In the \rtt\  plot (top right) we see an offset in the highest metallicity bin of 0.06 dex and 68$\%$ of the models span a range of 0.13 dex. For the three models with the lowest metallicity input the results of the stochastic models agree well with the non-stochastic ones. The highest offset of the relative metallicities is visible in the bottom left plot, where we present abundances calibrated with the \otnt\ method. Here all models have an offset, that increases towards increasing input metallicity, with a maximum offset of 0.11 dex for the \Z\ = 1 \Zsun\ models and a maximum spread of 0.20 dex of 68$\%$ of the models.

We conclude that the maximum relative offset and scatter are found for the solar metallicity models. At lower metallicity, only the \otnt\ calibration shows clear systematic offsets (at this \SFR ). Our choice of fixing the input metallicity to \Z\ = 1 \Zsun\ in the models of the previous and the next section ensures us of finding the maximum relative offsets among the metallicities considered here. 
\subsection{Influence of ionisation parameter}
\begin{figure*}
\captionsetup[subfigure]{labelformat=empty}
\centering
\subfloat[]{
  \includegraphics[width=93mm]{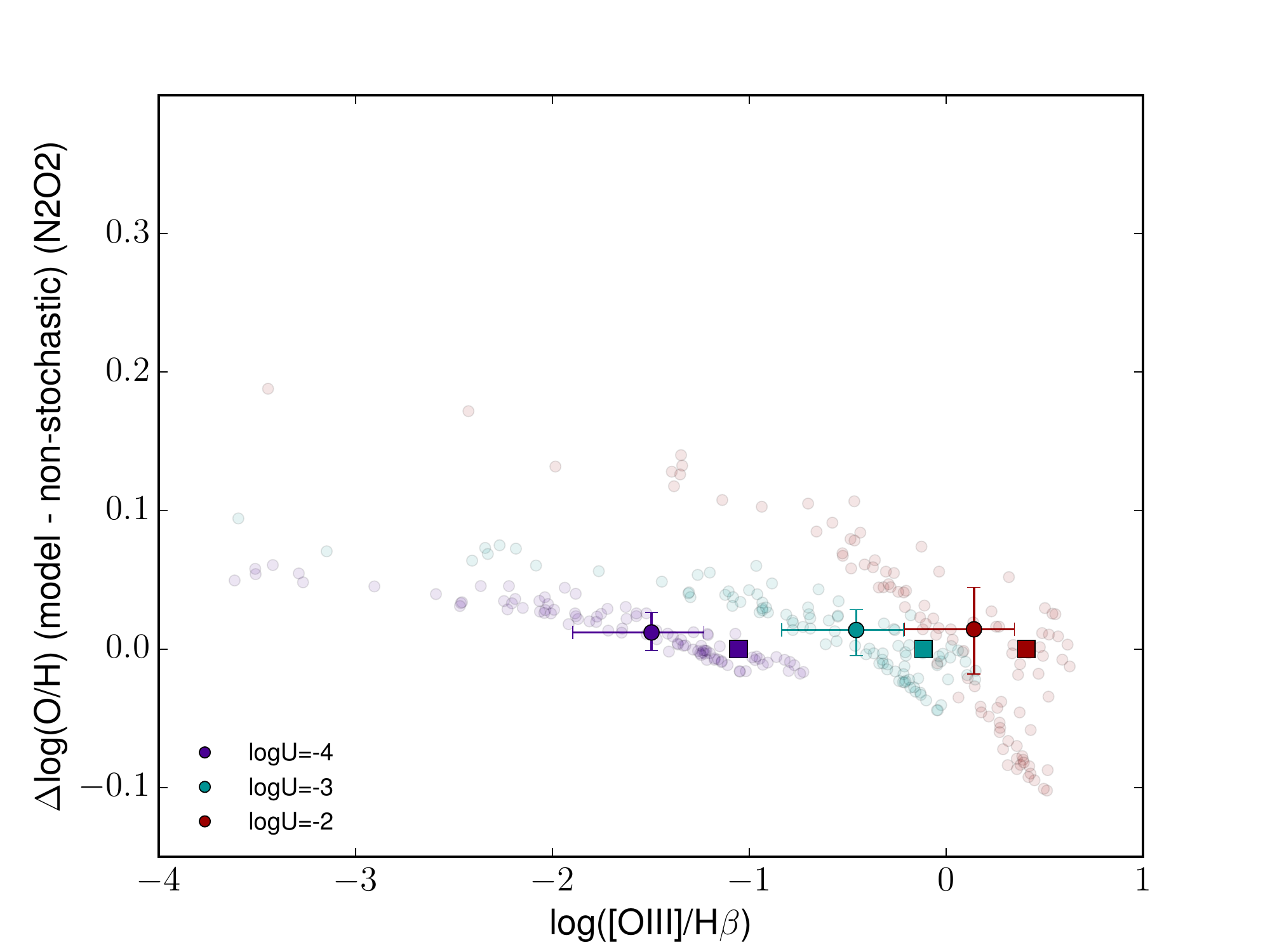}
}
\subfloat[]{
  \includegraphics[width=93mm]{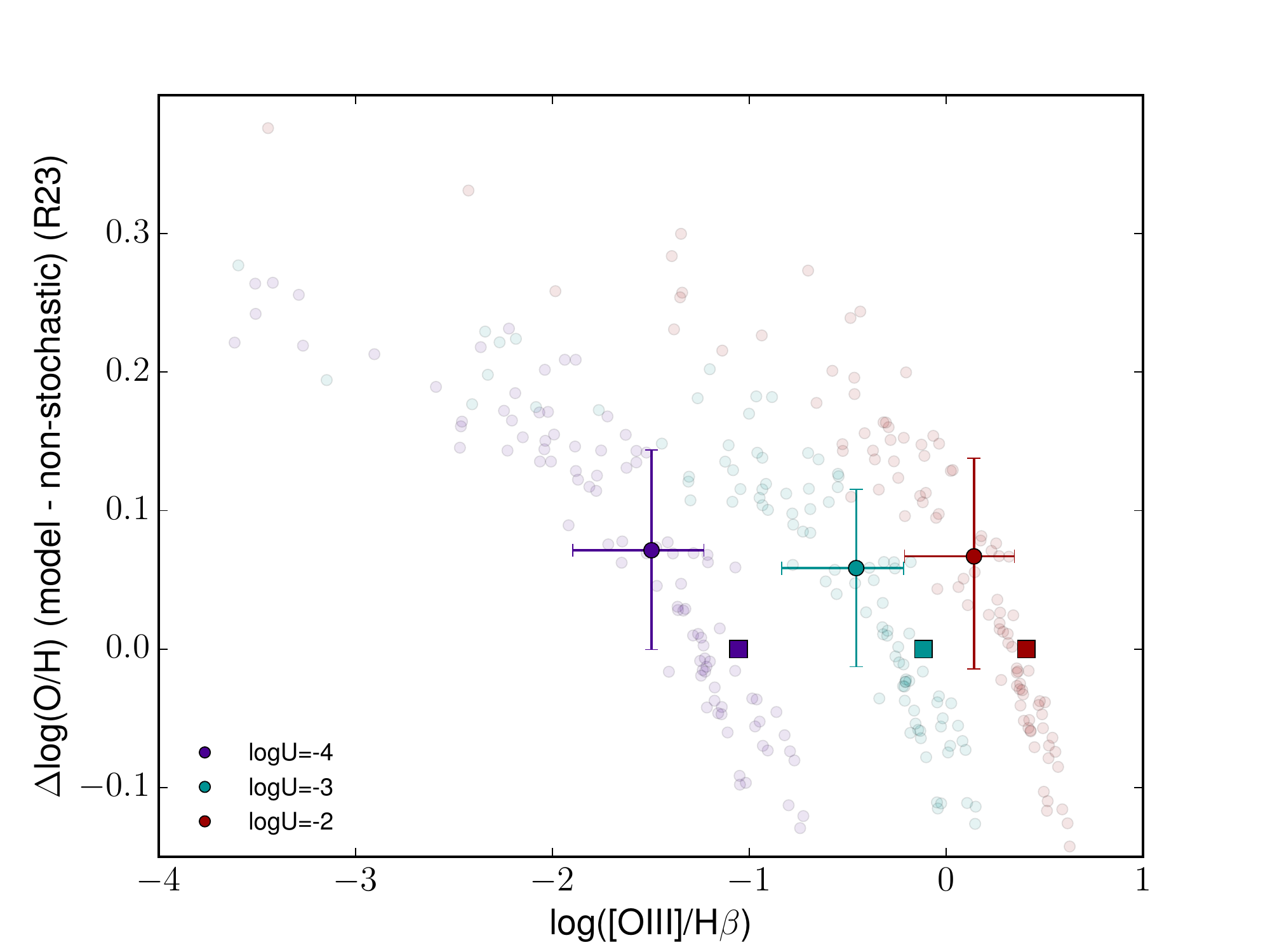}
}
\hspace{0mm}
\subfloat[]{
  \includegraphics[width=93mm]{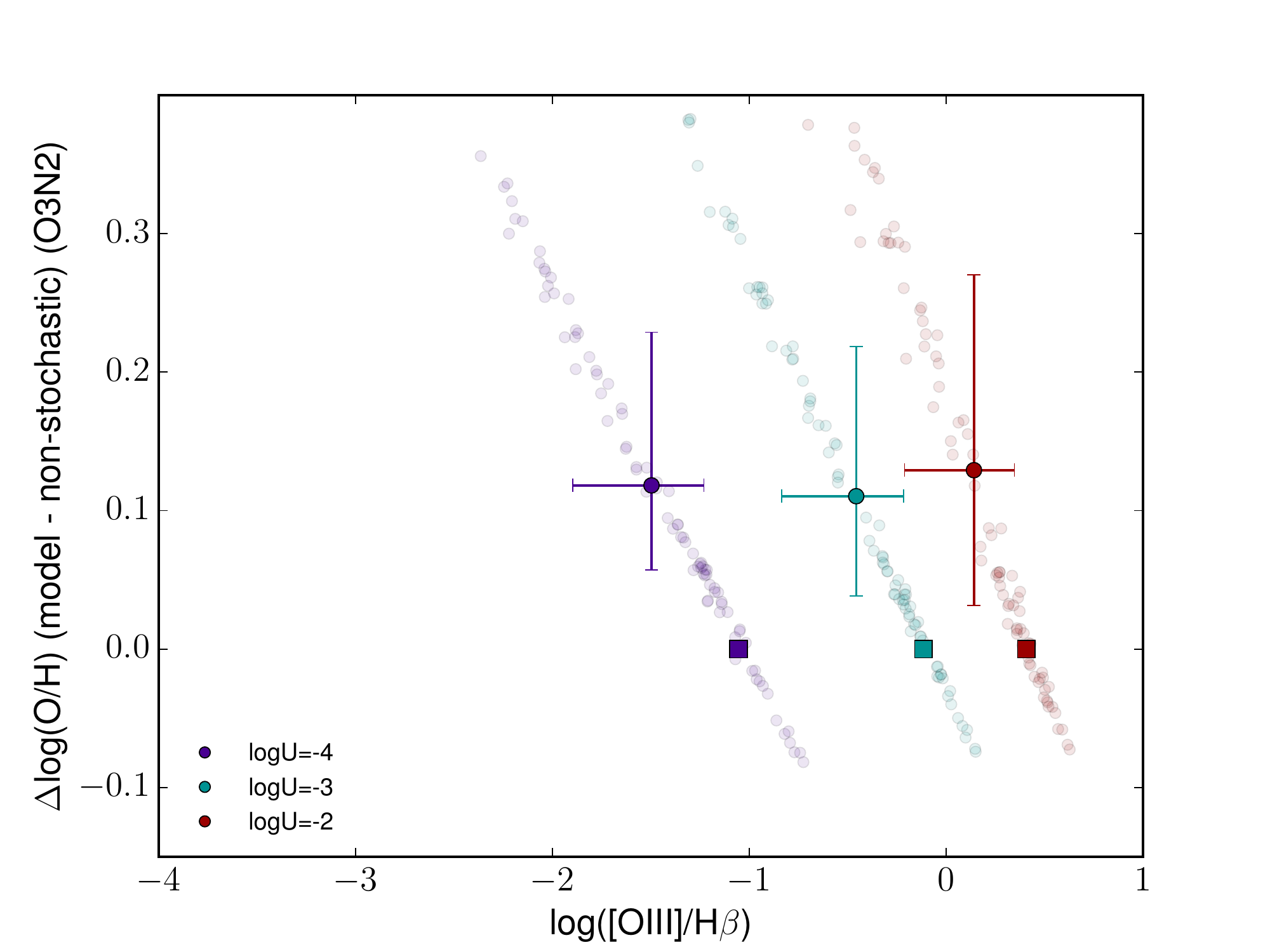}
}
\subfloat[]{
  \includegraphics[width=93mm]{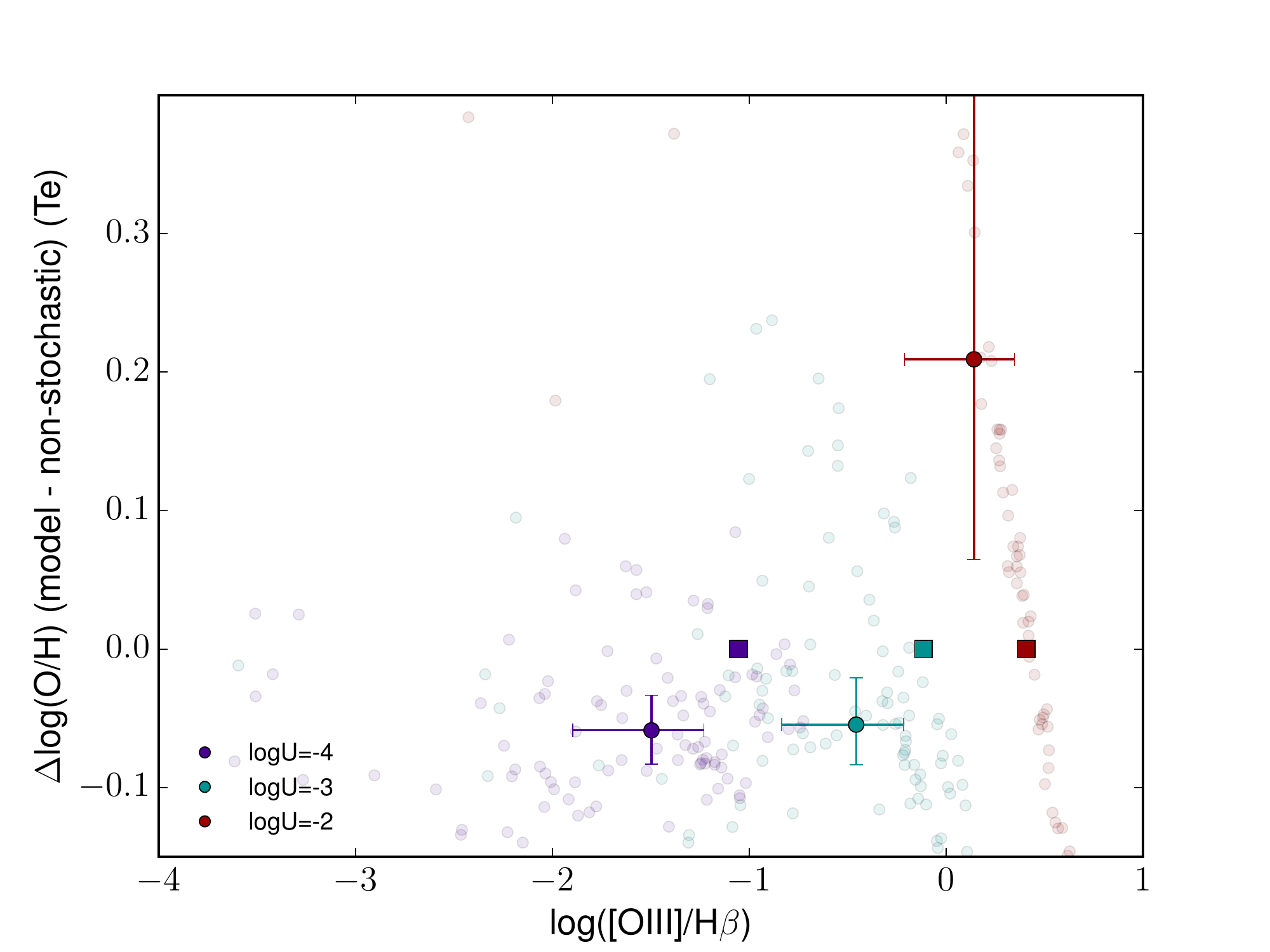}
}
\caption{Similar to Figures \ref{fig:resultsfr} and \ref{fig:resultmetal}, but the different colours match the ionisation parameter, with \logU\ = -4 (purple), \logU\ = -3 (cyan) and \logU\ = -2 (red). The input \Z\ of all models is 1 \Zsun and the \SFR = 0.001 \Msun/yr. Note that the axes range is different than that of the previous Figures. }
\label{fig:resultlogU}
\end{figure*}
In the previous section we adopted a fixed ionisation parameter of \logU\ = -3. In this section we will study the consequences of varying this at fixed other parameters. In Figure \ref{fig:resultlogU} we show the results of models with different ionisation parameter but fixed \SFR\ and \Z\ (\SFR\ = 0.001 \Msun/yr, \Z\ = 1 \Zsun). The purple points correspond to models with \logU\ = -4, the cyan with \logU\ = -3 and red with \logU\ = -2. We point out that these plots are presented with different axes compared to the plots in the previous sections, which needs to be taken into account when comparing them.  

Although the \oiii/\hb\ line ratios vary with \logU, there is minimal discrepancy between the relative offsets and scatter of the derived chemical abundances for our models with different ionisation parameters. The only exception is the results of the the \logU\ = -2 models that are calibrated using \Te\ (bottom right panel). In conclusion, the derived metallicities of our models are only weakly dependent on \logU\ and therefore adopting a fixed \logU\ to investigate the effects of a stochastic sampling, as we did in previous subsections, does not influence the results in general. The only exceptions are metallicities derived by \Te . Metallicities from models with \logU\ = -2  diverge from those from \logU\ = -3 and \logU\ = -4 models. 

\section{Discussion}
Let us now turn to a discussion of the underlying physical reason for these offsets. When an IMF is sampled stochastically with the framework used here, the consequence is that at low \SFR s we typically get an actual IMF somewhat depleted at the highest masses relative to a non-stochastic case. The consequence of this lack of massive stars is a lower production of \oiii\ relative to Balmer lines. If one interprets the line ratios with the assumptions of a fully populated, fixed IMF, the lower \oiii\ flux is interpreted as coming from a lower temperature gas and hence, at least at fixed $U$, a higher metallicity.  This then, is the reason for the sign of the observed offsets for \otnt\ and \rtt. We can apply the same reasoning to explain the offsets for \ntot , although this offset is smaller than those of the former two, since the energy that is necessary for the production of \nii\ is closer to that of \oii .

In the case of the \Te\ method, the situation is somewhat different. Since the energy that is required to produce the auroral emission line \oiii{\ensuremath{\lambda}4363} is higher compared to \oiii{\ensuremath{\lambda}4959} and \oiii{\ensuremath{\lambda}5007}, this causes the \oiii{\ensuremath{\lambda\lambda}4959,5007}/\oiii{\ensuremath{\lambda}4363} ratio to increase in situations with an underpopulation of massive stars, which results in a decrease of the observed \Te (\oiii). However, metallicities that are calibrated by this method are very robust to the effects of stochasticity for models with \Z\ < 1 \Zsun. We therefore expect that stochastic effects on the \oiii\ lines are canceled out by these effects on O$^{+}$/H$^{+}$ and O$^{++}$/H$^{+}$. However, the scatter on the relative metallicity of our \Z\ = 1 \Zsun\ models is larger than that of the sub-solar models.

By default we calculated \Te (\oii) from \Te (\oiii) following the relation of \citet{2006A&A...448..955I}. To test whether this influences the metallicity calibrations of the solar metallicity models,  we also derive the electron temperature in the \oii\ region from the \oii{\ensuremath{\lambda}3727}/\oii{\ensuremath{\lambda}7325} line ratio.   We find that the 1-$\sigma$ scatter of the relative metallicity decreases from 0.10 dex to 0.038 dex if we use the directly measured \Te (\oii) in stead of the \Te (\oii) from the relation with \Te (\oiii). For the solar metallicity models, most of the scatter of the relative metallicity thus originates from the \Te (\oii) estimations, that are less well constrained by the \Te (\oiii) - \Te (\oii) relation in this metallicity regime. 

However, while there are clearly better options for measuring gas-phase abundances at low \SFR s, these are not always available. One might therefore ask whether a correction method can be found to correct for the effects of stochasticity and we turn to this next.
\subsection{Tracing a stochastic IMF}
\begin{figure*}
\captionsetup[subfigure]{labelformat=empty}
\centering
\subfloat[]{
  \includegraphics[width=93mm]{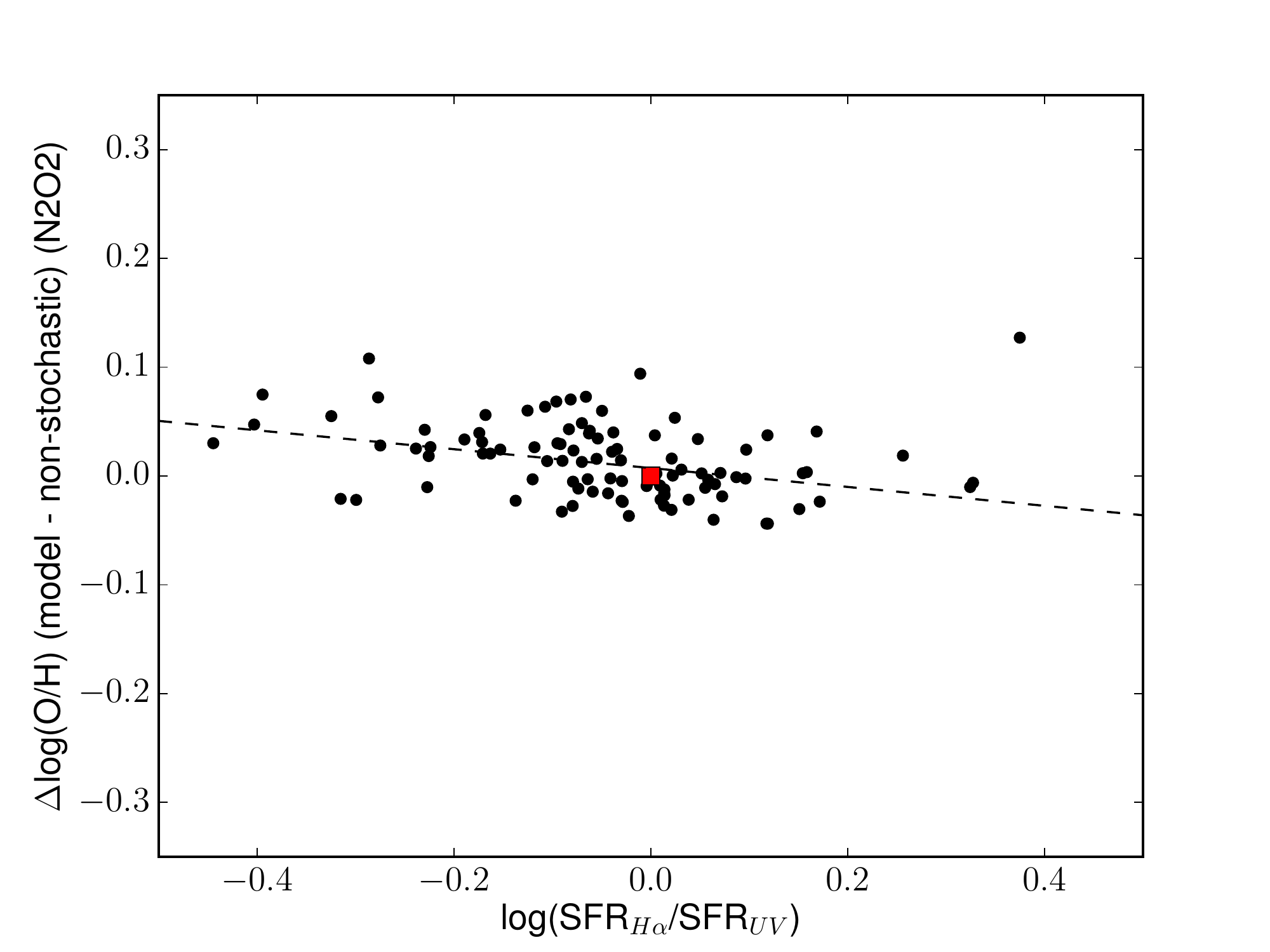}
}
\subfloat[]{
  \includegraphics[width=93mm]{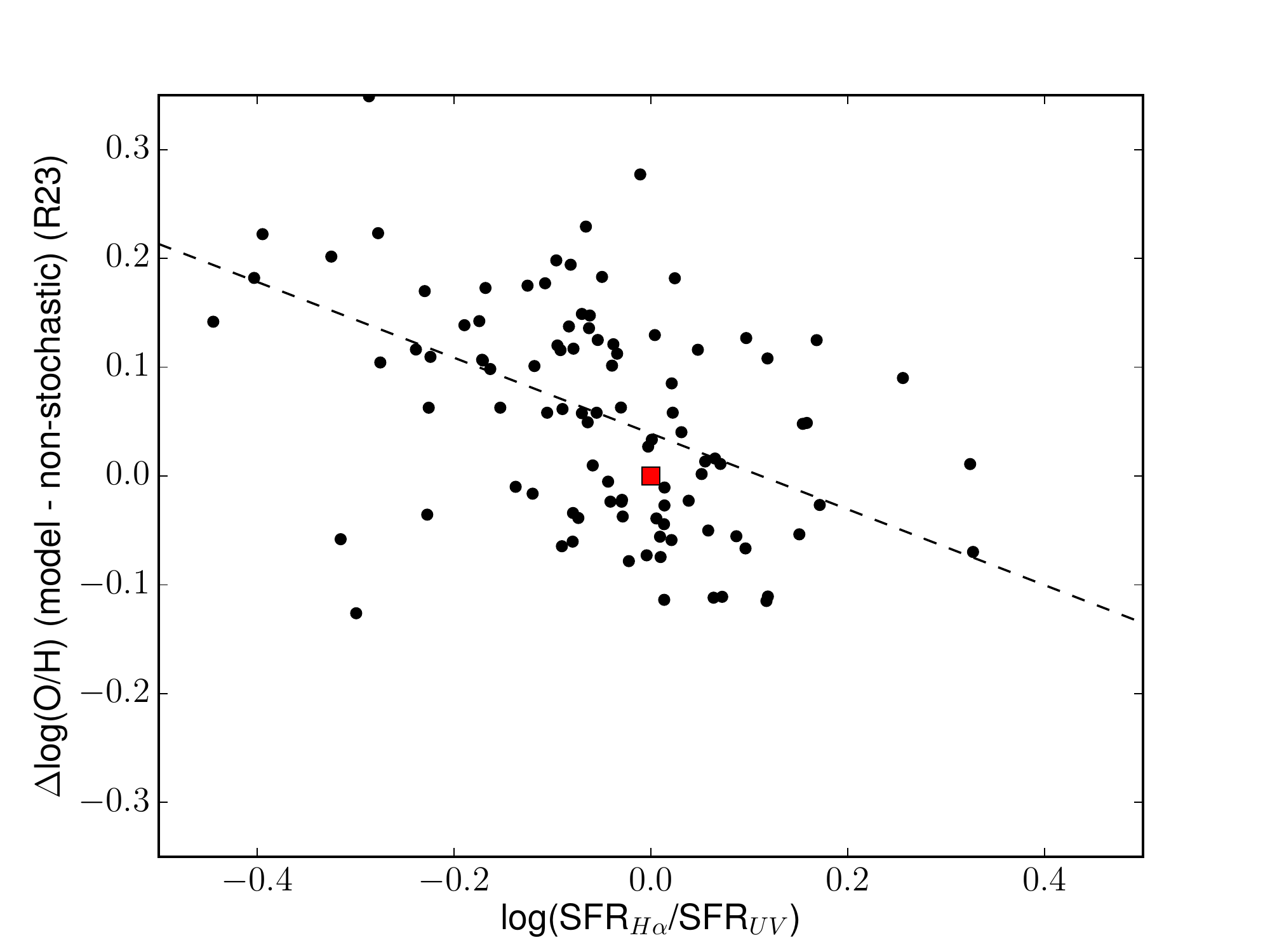}
}
\hspace{0mm}
\subfloat[]{
  \includegraphics[width=93mm]{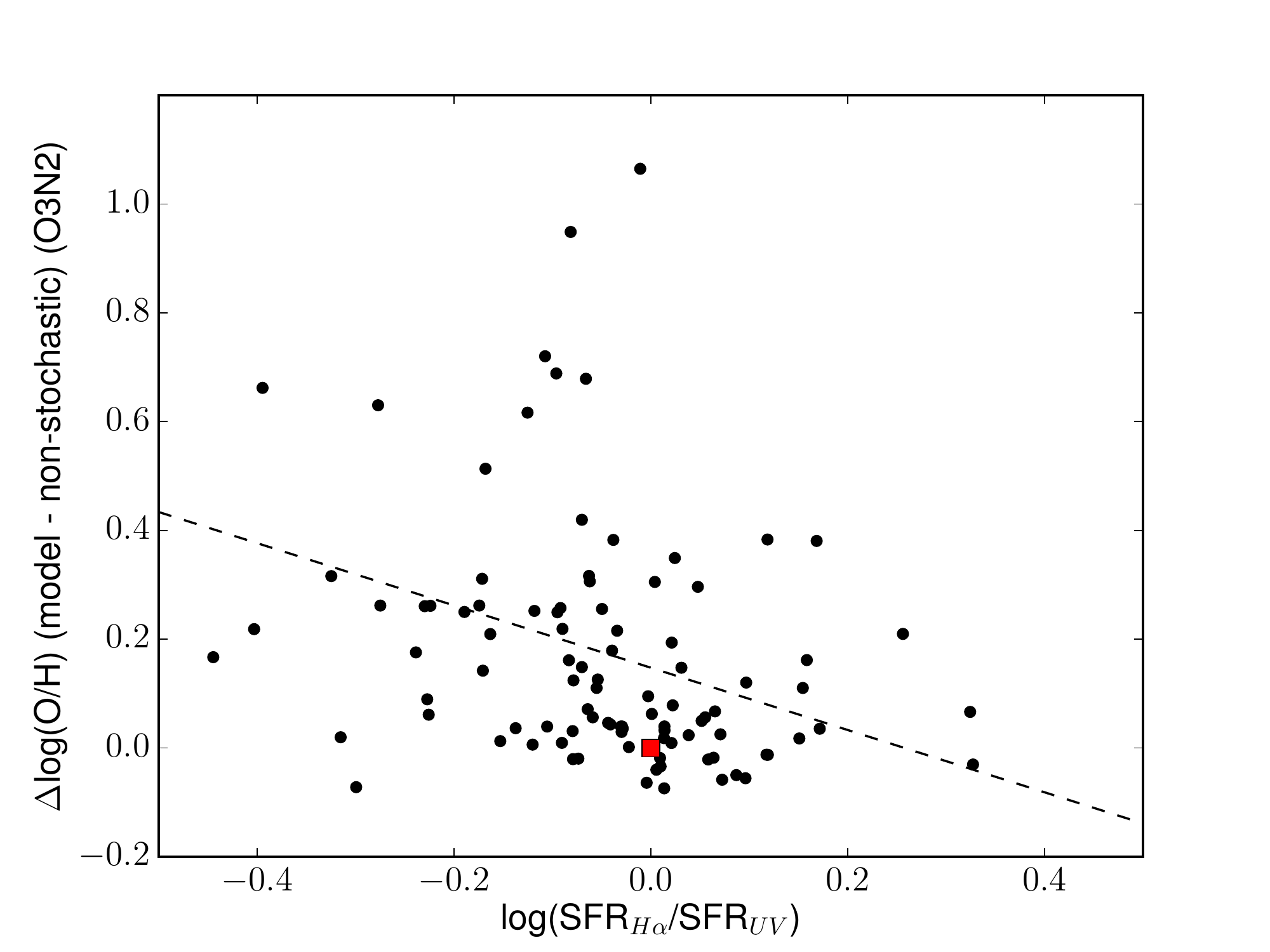}
}
\subfloat[]{
  \includegraphics[width=93mm]{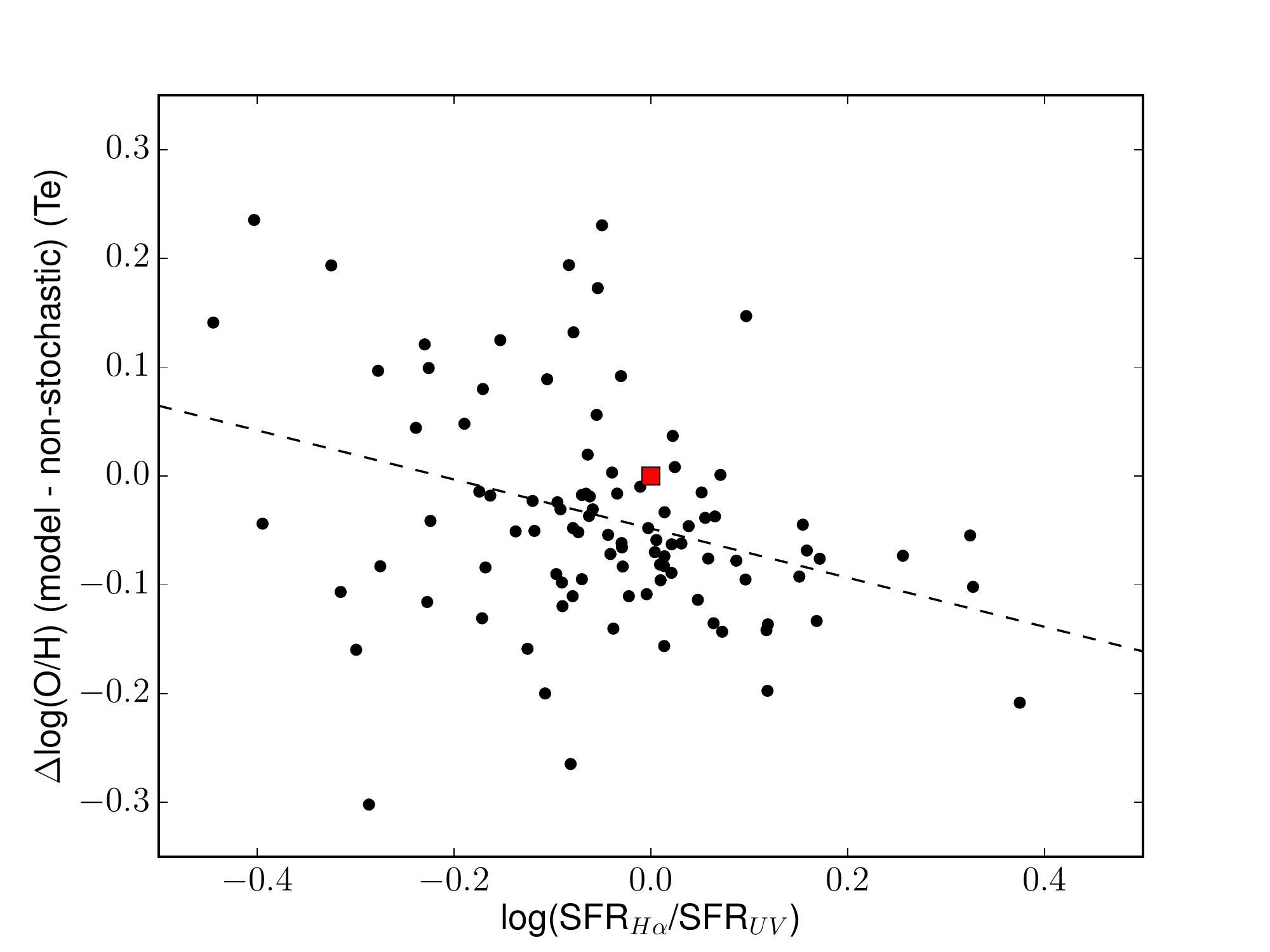}
}
\caption{The log(O/H) of the stochastic models relative to the log(O/H) of a typical non-stochastic model, versus the ratio of the \SFR\ derived by \ha\ over the \SFR\ that is calculated from the UV luminosity. The black dots represent the results of our models with \Z\ = 1 \Zsun, \SFR\ = 0.001 \Msun/yr and \logU\ = -3. The non-stochastic model with the same properties is shown with the red square. We point out that the y-axis of the \otnt\ plot deviates from the y-axis of the other plots. The dashed line is a fit of the data from $\chi^2$ minimization.}
\label{fig:sfrhauv}
\end{figure*}
The first signs of a possible stochastically sampled IMF of low star-forming galaxies originate from inconsistent \SFR\ measurements. This presents us with a possible way to identify and mitigate stochastic effects in abundance determinations, since the models for which we found the effects to be large, above, are also those where one expects large variation in \SFRHa/\SFRUV\ (e.g. \citealt{2009ApJ...706..599L,2009ApJ...695..765M})\\
\\
In Figure \ref{fig:sfrhauv} we present the relative abundances ($\Delta$ log(O/H)) with the \ha-to-UV \SFR\ of our models. We normalize the \SFRs\ in such a way that the \ha\ \SFR\ equals the input \SFR\ (0.001 \Msun/yr in this case), because the ionising energy from the stellar light is scaled to the amount of hydrogen ionising photons. To test whether the offset of the relative metallicity is related to the \SFRHa/\SFRUV\ ratio, we performed a $\chi^2$ minimization fit to the data and a Spearman rank-order correlation test to assess the quality of the fit. The details of this are presented in Table \ref{tab:fit_results}. 
\begin{table}
	\centering
	\begin{tabular}{ccc} 
		\hline
		\hline
		 Calibration & $\rho$ & slope\\
		\hline
		\ntot & -0.43 & -0.09 \\
		\rtt & -0.45 & -0.35 \\
		\otnt & -0.38 & -0.57 \\
		\Te & -0.29 & -0.23 \\
		\hline
		\hline
    	\end{tabular}
	\caption{Results of the $\chi^2$ minimization fits of our results in Figure  \ref{fig:sfrhauv}. $\rho$ is the correlation coefficient derived from a Spearman rank-order correlation test.}
        \label{tab:fit_results}
\end{table}
While there appears to be a weak trend, the Spearman rank-order coefficient, $\rho$, is large, indicating a weak correlation ($\rho$ is between 0 (no correlation) and -1 (optimal fit) for a fit with a negative slope).  Thus while the \SFRHa/\SFRUV\ ratio might indicate that the IMF is stochastically sampled, this is not directly related to the amount of offset in the derived oxygen abundances, caused by stochastic sampling. Therefore, corrections for stochastic sampling are hard to perform based on the \SFRHa/\SFRUV\ ratio. We thus conclude that performing metallicity calibrations of galaxies with \SFR\ would benefit from a combination of several calibrators.

\subsection{Other variations of the IMF}
As we mentioned in the introduction, stochastic sampling is not the only explanation for the discrepancy in observed \SFRHa\ and \SFRUV . We will here compare the impact of truncated IMF, a variation of the slope of the IMF, and bursty star formation on metallicity calibrations to those that we demonstrated for stochastic sampling. 

If we had applied our analysis on models were stellar masses are sampled from an IGIMF, we expect a comparable offset in metallicity to the ones we obtained for the stochastic models. Since there will be no situation with higher abundances of massive stars as we observed for the stochastic models, there will be no scatter in derived metallicities in the opposite direction of the offset (see also the results in \citealt{2011ApJ...741L..26F}). This will result in narrower dispersion of metallicities, that is focussed around the offset that we determined for the stochastic models.

When stellar masses are distributed by a top-light IMF, the obtained metallicities will be consistent with those from the IGIMF, with the observed offset depending on the slope of the IMF. The opposite is true for situations where stellar masses are drawn following a top-heavy IMF, e.g. the relative offset will be of opposite sign. 

Last, we discuss the situation where the star formation history is bursty, which is often the case for low-mass galaxies, as we mentioned in the introduction. The consequence of a rapidly changing \SFR\ on metallicity derivations is not straightforward. Generally, we argue that during or directly after a burst, the massive star distribution does not deviate significantly from a 'normal' IMF, but in the periods between two bursts, this will do so. Therefore, depending on the time that we observe a galaxy with bursty star formation, the derived metallicity is either not affected by a lack of massive stars, or it is, and this translates back to a similar offset that we observed for our models. 

\section{Conclusion}
The main results of this paper are summarised by the following points: 
\begin{itemize}
\item A stochastically sampled IMF causes variation in the ratios between emission lines, and this effect increases with decreasing \SFR\ and is visible for models with any input parameter and ionisation potential. The scatter of the line ratio is larger if the energy difference between emission lines is higher. For models with \SFR\ = 0.0001 \Msun/yr, a significant part of the \oiii\ emission lines are below a detection limit of $10^{-18}$ ergs/s/cm$^2$, that we set as a reference. The line ratios for the \Z\ = \Zsun\ , \SFR\ = 0.1 \Msun/yr and $\log U$ = -2 and -3, are in good agreement with those from observed SDSS galaxies. 
\item As an effect, the determined metallicity significantly changes for calibrators that are based on emission line ratios of lines with a widely different ionisation potential. The estimated abundances are scattered towards higher metallicities when the \ntot , \rtt\ and \otnt\ calibrators are used, because of the relatively lower number of high energy photons available to doubly ionised oxygen. For models with  \Z\ = \Zsun\ the \Te\ method provides an underestimation of the metallicity, but this method is robust for lower input metallicities. 
\item The induced scatter in determined abundances is most prominent for our models with \SFR\ = 0.0001 and 0.001 \Msun/yr and \Z\ = 1 \Zsun, and is fairly independent of ionisation parameter.  
\item We found relations between the scatter in our metallicity calibrations and the measured \SFRHa/\SFRUV\ for \ntot , \rtt , \otnt\ and \Te\ estimations. Although the correlations are weak, they provide a first tool to correct metallicity calibrations in observational studies to low star-forming galaxies.  
\end{itemize}
\section*{Acknowledgements}
We thank the referee for useful comments and suggestions. We would also like to thank Madusha Gunawardhana for providing useful discussions for this work. This work is supported by Funda{\c c}{\~a}o para a Ci{\^e}ncia e a Tecnologia (FCT) through national funds (UID/FIS/04434/2013) and by FEDER through COMPETE2020 (POCI-01-0145-FEDER-007672). JB is supported by FCT through Investigador FCT contract IF/01654/2014/CP1215/CT0003.




\bibliographystyle{mnras}
\bibliography{myrefs} 

\bsp	
\label{lastpage}
\end{document}